\documentclass[11pt]{article}
\usepackage{graphicx}
\usepackage{array}
\usepackage{comment}
\usepackage[hyphens]{url}
\usepackage{cite}

\begin{document}
\title{RevDedup: A Reverse Deduplication Storage System Optimized for Reads to
	Latest Backups}
\author{Chun-Ho Ng and Patrick P. C. Lee\\
The Chinese University of Hong Kong, Hong Kong\\
Technical Report\\
\{chng,pclee\}@cse.cuhk.edu.hk}

\maketitle

\begin{abstract}
Scaling up the backup storage for an ever-increasing volume of virtual machine
(VM) images is a critical issue in virtualization environments.  While
deduplication is known to effectively eliminate duplicates for VM image
storage, it also introduces fragmentation that will degrade read performance. 
We propose {\em RevDedup}, a deduplication system that optimizes reads
to latest VM image backups using an idea called reverse deduplication.  In
contrast with conventional deduplication that removes duplicates from new
data, RevDedup removes duplicates from old data, thereby shifting 
fragmentation to old data while keeping the layout of new data as
sequential as possible.  We evaluate our RevDedup prototype using
microbenchmark and real-world workloads. For a 12-week span of real-world VM
images from 160 users, RevDedup achieves high deduplication efficiency with
around 97\% of saving, and high backup and read throughput on the order of
1GB/s.  RevDedup also incurs small metadata overhead in backup/read operations. 
\end{abstract}

\section{Introduction}
\label{sec:introduction}

Many enterprises today adopt virtualization technologies to run a large number
of virtual machines (VMs) on a small group of physical hosts.
%Virtualization facilitates the administration of computational and storage
%resources with reduced hardware footprints.  
For disaster recovery, it is necessary to preserve user data and any operating
system updates made to a VM.  Conventional approaches schedule backups for
each VM disk image and keep different versions of each VM backup, so that
administrators can restore any previous recovery checkpoint.
%VM versioning also allows users to save work-in-progress jobs in VMs and
%restart the VM states on different physical hosts to make development
%flexible.  
Today's backup solutions are mainly based on disk-based storage,
which has better I/O performance than traditional tape-based storage.
However, each VM image typically contains several gigabytes of data.
Even though the cost of disk-based storage is low nowadays, in the face of a
large volume of VMs and a large volume of versions associated with each
VM, scaling up the backup storage for VM images still remains a critical
deployment issue. 

Deduplication improves storage efficiency by eliminating redundant data.
Instead of storing multiple copies of data blocks that have identical content,
a deduplication system stores only one copy of identical blocks, while other
blocks refer to the copy via smaller-size references.  Deduplication is mainly
studied in content-addressable backup systems (see \S\ref{subsec:related}). It
is also shown to provide space-efficient VM image storage given that VM images
have significant content similarities \cite{jayaram11,jin09,liguori08,nath06}.  
%backup (see Section~\ref{sec:related}).  It is also shown to
%\cite{bhagwat09,guo11,kruus10,lillibridge09,quinlan02,rhea08,you05,zhu08}. 

Most existing deduplication studies focus on optimizing storage efficiency and
write (or backup) performance.  However, one drawback of deduplication is that
it introduces {\em fragmentation}, since some blocks of a file may now refer
to other identical blocks of a different file.  Hence, accessing a file is no
longer sequential as in ordinary file systems without deduplication, but
instead requires additional disk seeks to the identical blocks being
referenced. This significantly degrades read performance.  On the other hand,
we believe that achieving high read throughput is necessary in any backup
system.  For instance, a fast restore operation can minimize the system
downtime during disaster recovery.  Also, enabling high read performance makes
new applications feasible.  For example, administrators can retrieve recently
archived VM images to conduct forensic analysis. 

In this work, we explore the use of deduplication for VM image backup storage
on a disk-based backend.  Our goal is to maintain high read throughput as in
ordinary file systems without deduplication, while maintaining high write
performance and high storage efficiency as in existing deduplication systems.
%We focus on {\em inline} deduplication, meaning that deduplication is
%performed on the write path.  
In practice, users are more likely to access
more recent data.  Our key insight is that traditional deduplication
systems check if new blocks can be represented by any already stored blocks
with identical contents.  Thus, the fragmentation problem of the latest backup
is the most severe since its blocks are scattered across all the prior
backups.  To mitigate fragmentation in newer backups, we propose to
do the opposite, and check if any already stored blocks can be represented by
the new blocks to be written.  We remove any duplicate blocks that are already
stored so as to reclaim storage, and refer them to the new blocks.  This
shifts the fragmentation problem to the older backups, while keeping the
storage layout of the newer backups as sequential as possible.  We call this
{\em reverse deduplication}, which is the core component of our deduplication
design.

To this end, we propose {\em RevDedup}, a deduplication system for 
VM image backup storage.  RevDedup exploits content similarities of VM images
using a hybrid of inline and out-of-order deduplication
approaches.  It applies coarse-grained global deduplication (inline) to
different VMs and removes any duplicates on the write path, and further
applies fine-grained reverse deduplication (out-of-order) to different backup
versions of the same VM and removes any duplicates from old backup versions.
We propose a configurable, threshold-based block removal mechanism that
combines {\em hole-punching} \cite{lwn} to remove duplicate blocks of old
backup versions and {\em segment compaction} to compact data segments without
duplicate blocks to reclaim contiguous space.   

We implement RevDedup based on a client-server model, which allows multiple
clients to submit changes of VM images to a storage server.  We experiment
our RevDedup prototype on a RAID disk array using microbenchmark and real-world
workloads.  In particular, we collected a dataset of weekly VM image snapshots
for 160 university students in a computer science programming course over a
12-week span.  We show via this dataset that RevDedup achieves (i) high
deduplication efficiency with around 97\% of saving, (ii) high write
throughput at 4-7GB/s, and (iii) high read throughput for the
latest backup at 1.2-1.7GB/s.  We also show that conventional deduplication
experiences throughput drop when retrieving newer backups.  Finally, we show
that RevDedup incurs small metadata overhead in backup/read operations when it
operates on a VM backup with a large number of versions.  To our knowledge,
this is the first work that provides prototype implementation of a
deduplication storage system that is optimized for reads to latest backups.

% We compare RevDedup with ZFS on Linux \cite{zfsonlinux}, an open-source
% deduplication file system originated from Sun Microsytems. We show that
% RevDedup outperforms ZFS in both read and write operations.

The rest of the paper proceeds as follows.  
In \S\ref{sec:background}, we discuss the fragmentation problem in
deduplication and review related work.  
In \S\ref{sec:design}, we describe the design and implementation of
RevDedup.
In \S\ref{sec:evaluation}, we present experimental results.
Finally, in \S\ref{sec:conclusion}, we conclude the paper.

\section{Background and Motivation}
\label{sec:background}

Deduplication is a well-known technique for exploiting content similarities
and eliminating the storage of redundant data.  Typical deduplication systems
divide a backup stream into {\em blocks}, and use {\em fingerprints} to
identify blocks and check if blocks can be deduplicated.  A fingerprint is
computed by a cryptographic hash (e.g., MD5, SHA-1) of the content of a block.
Two blocks are said to be identical if their fingerprints are the same. We
assume that the probability that two different blocks have fingerprint
collisions is negligible \cite{black06}.

\subsection{Fragmentation} 

\begin{figure*}[t]
\centering
\begin{minipage}{0.205\linewidth}
\centering
\includegraphics[width=\linewidth]{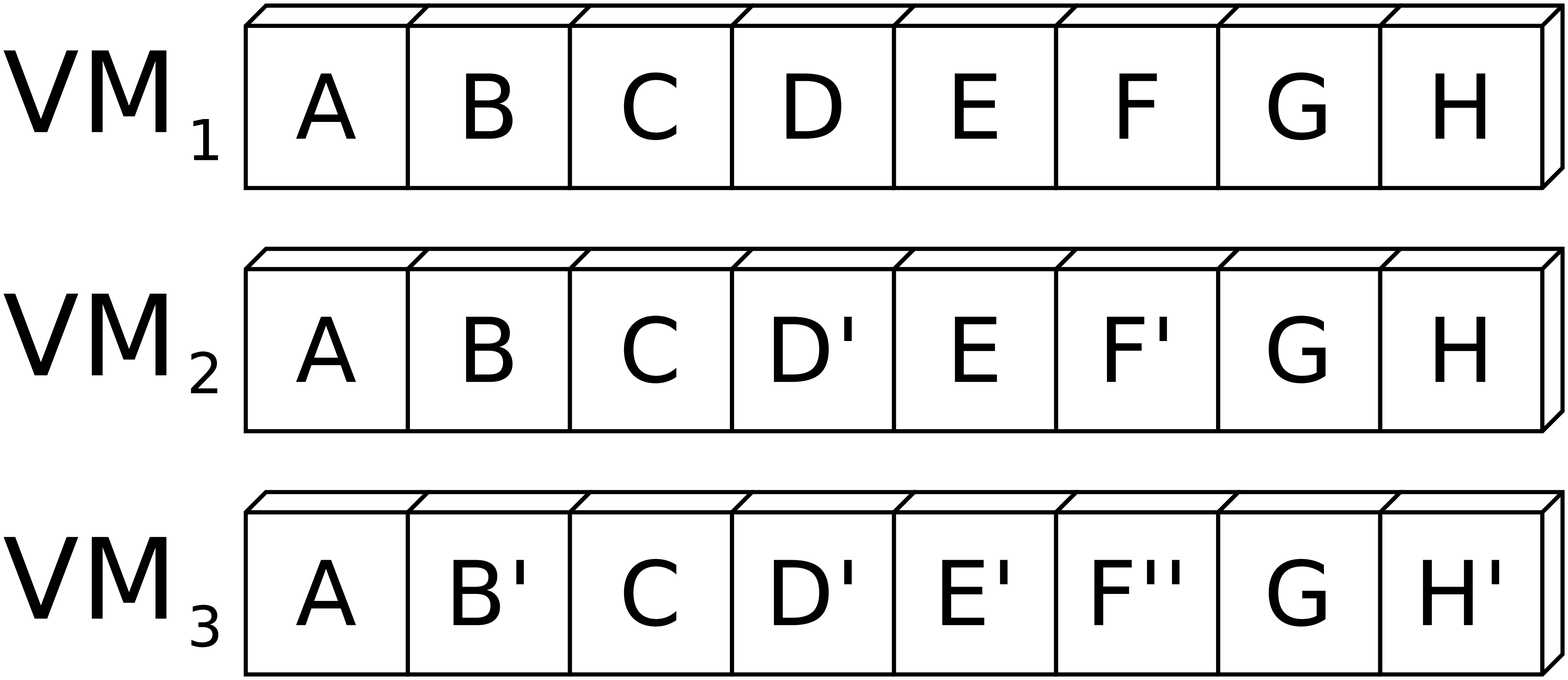}
\mbox{\small (a) Backups}
\end{minipage}%
\hspace{3pt}
\begin{minipage}{0.36\linewidth}
\centering
\includegraphics[width=\linewidth]{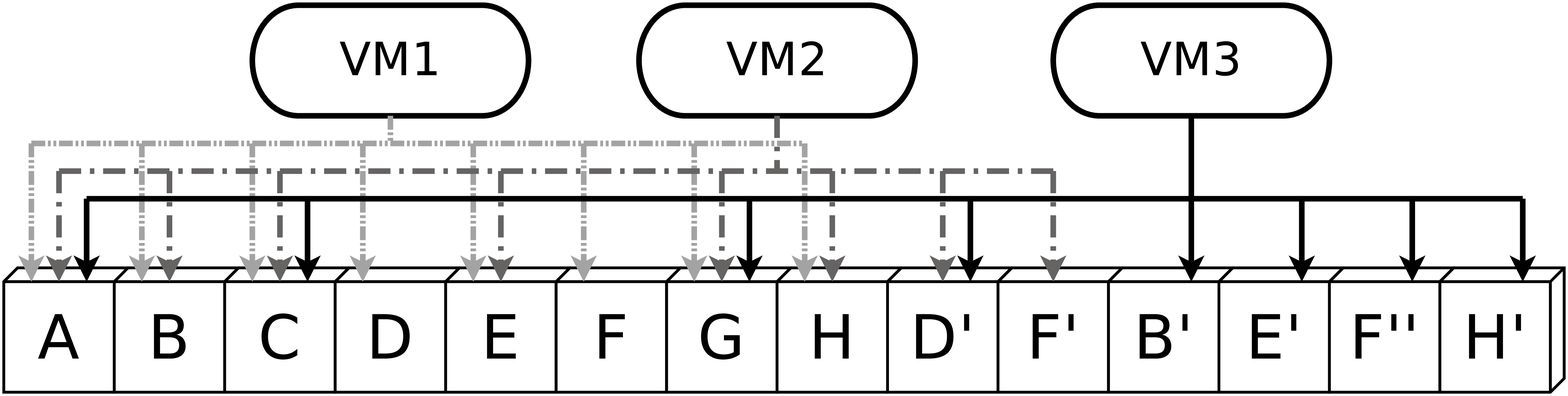}
\mbox{\small (b) Conventional deduplication}
\end{minipage}%
\hspace{3pt}
\begin{minipage}{0.39\linewidth}
\centering
\includegraphics[width=\linewidth]{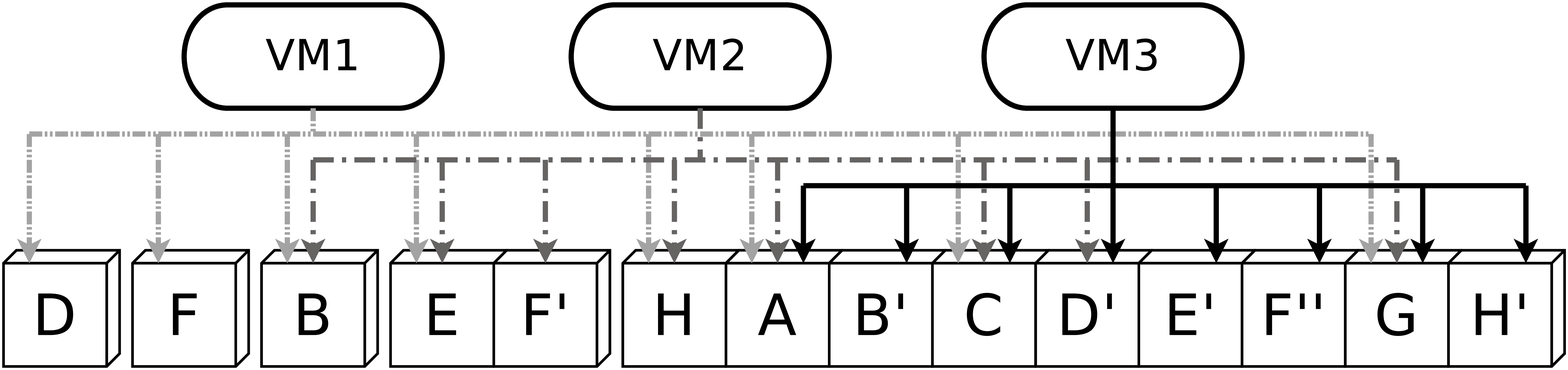}
\mbox{\small (c) RevDedup}
\end{minipage}%
\vspace{-3pt}
\caption{An example of how conventional deduplication and RevDedup place data
on disk.}
\label{fig:eg_trad}
\end{figure*}

Most deduplication systems suffer from the inherent fragmentation problem,
which has also been addressed in prior work
\cite{kaczmarczyk12,nam12,rhea08,srinivasan12}.  We now illustrate the
fragmentation problem using a simple example.  Figure~\ref{fig:eg_trad}(a)
shows that a deduplication system is about to write, in order, three snapshots
of a VM, denoted by VM$_1$, VM$_2$, and VM$_3$.  We assume that each VM image
has eight blocks, and that the system has no data initially.
Figure~\ref{fig:eg_trad}(b) shows how a conventional
deduplication system writes data.  First, the system writes VM$_1$ with unique
blocks {\sf A} to {\sf H}.  Given that all blocks are new, the system will
sequentially write all of them to disk.  Next, the system writes VM$_2$, in
which some of the blocks are identical to those of VM$_1$. Then the system
stores only the references that refer to those identical blocks,
and appends the unique blocks {\sf D'} and {\sf F'} to the end of the last
write position.  The same approach applies when the system writes VM$_3$, and
it only writes the unique blocks {\sf E'}, {\sf F''}, and {\sf H'} to the end
of the last write position.  

%
%\begin{figure}[t]
%\centering
%\includegraphics[height=\linewidth,angle=270]{figs/eg_timededup}
%\caption{An example of how a RevDedup places data on disk.}
%\label{fig:eg_timededup}
%\end{figure}

From Figure~\ref{fig:eg_trad}(b), we see that the blocks of the latest written
image VM$_3$ are re-ordered and randomly scattered across the previously
written images VM$_1$ and VM$_2$.  Reading VM$_3$ will generate disk seeks and
see degraded performance.  We can perform simple calculation to understand the
degradation.  Consider a generic 7200~RPM 3.5-inch SATA harddisk with an
average seek time of 8.5ms \cite{seagate}.  With only 60 disk seeks per second
on the read path, the read throughput can drop by at least 50\% compared to
the sequential read (which we assume has negligible seek time). 

Maintaining high read performance for deduplication backup systems is
necessary for minimizing the system downtime during disaster recovery (see
\S\ref{sec:introduction}).  We note that the latest backup contains the
``hot'' data and is expected to be more likely read in practice, while older
backups contain ``cold'' data that usually serves the compliance purpose and
is less likely read.   This guides our design of RevDedup.
Figure~\ref{fig:eg_trad}(c) shows the disk layout for the previous example
when RevDedup is used.  We allow the blocks of VM$_3$ (i.e., the latest
backup) to be sequentially placed on disk.  Also, VM$_2$ is less fragmented
than VM$_1$, in the sense that the blocks of VM$_2$ are less spread out on
disk than those of VM$_1$.  Thus, the newer a backup is, the less
fragmentation overhead the backup will experience.  We explain how we achieve
this property in \S\ref{sec:design}. 

\subsection{Related Work}
\label{subsec:related}

Deduplication has been widely used in backup applications.  We review related
work on deduplication storage. 

{\bf Deduplication for backup storage.} Most existing deduplication studies
for backup storage focus on optimizing fingerprint indexing to achieve high
backup performance.  Deduplication is first proposed in Venti \cite{quinlan02}
for data backup in content-addressable storage systems.  DDFS \cite{zhu08} and
Foundation \cite{rhea08} use Bloom-filter-based \cite{bloom70} indexing
structures to minimize memory usage.  DDFS further exploits spatial locality
to cache the fingerprints of blocks that are likely written later.  Other
studies \cite{lillibridge09,bhagwat09,kruus10,guo11,xia11} exploit workload
characteristics to reduce memory usage for indexing. 
 
%keeps only small portions of index samples in memory by exploiting chunk
%locality, such that the chunk orders of different backup streams are likely
%preserved.  
%exploits file similarity of backup streams, and keeps only a 
%representative fingerprint per file in memory.  A similarity signature scheme
%is proposed in \cite{aronovich09} to identify similar large-size chunks for
%deduplication.  Bimodal \cite{kruus10} reduces memory usage via different
%granularities of backup regions.  ChunkStash \cite{debnath10} and Dedupv1
%\cite{meister10} use solid state drives (SSDs) to store fingerprints.  The
%fast random I/Os of SSDs enable high-performance deduplication compared to the
%solutions based on memory and disk only.  Guo {\em et al.} \cite{guo11}
%propose sampled indexing and trades storage efficiency for less memory usage,
%and pipelined implementation to increase the backup throughput.  We point
%out that \cite{guo11} also experiments the read performance using their
%deduplication design, but it does not provide any solution that
%specifically optimizes the read performance.  Silo \cite{xia11} improves
%Extreme Binning \cite{bhagwat09} by packing small files into larger segments.

The above studies aim to achieve high write throughput, low memory usage, and
high storage efficiency, but put limited emphasis on read performance.  
%Guo {\em et al.} \cite{guo11} show via experiments that their deduplication
%prototype achieves around 40\% of the raw disk throughput in reads, but they
%do not provide any explicit solution that optimizes read performance.  
One closely related work to ours is by Kaczmarczyk {\em et al.}
\cite{kaczmarczyk12}, who also improve read performance for latest backups in
deduplication.  Their system selectively rewrites deduplicated data to disk to
mitigate fragmentation, and removes old rewritten
blocks in the background.  However, they consider deduplication for different
versions of a single backup only, while we enable global deduplication across
multiple VMs.  Nam {\em et al.} \cite{nam12} propose a system that measures
the fragmentation impact given the input workload and activates selective
deduplication on demand.  Lillibridge {\em et al.} \cite{lillibridge13} use
the container capping and forward assembly area techniques to improve the
restore performance.  Unlike the previous studies
\cite{kaczmarczyk12,nam12,lillibridge13} that aim to remove duplicates of new
data, we use a completely different design by removing duplicates of old data
to maintain high deduplication efficiency and inherently making older (newer)
backups more (less) fragmented. 
Note that the studies \cite{kaczmarczyk12,nam12,lillibridge13} only conduct
simulation-based evaluations, while we implement a prototype to experiment
the actual I/O throughput.

{\bf Distributed deduplication.} DeDe \cite{clements09} targets a storage area
network (SAN) connecting multiple client hosts that run VM instances.  It
performs out-of-order deduplication in the hosts to minimize the
synchronization overhead.  HYDRAstor \cite{dubnicki09} and its
successor HydraFS \cite{ungureanu10} are distributed deduplication systems
with multiple storage nodes. Our work focuses on a single storage backend. 

{\bf Deduplication for primary storage.} Several file systems (e.g.,
\cite{elshimi12,lessfs,ng11,opendedup,zfs}) deploy inline deduplication for
primary storage. In particular, iDedup \cite{srinivasan12} is a primary,
inline deduplication system that optimizes read performance.  It
applies deduplication to chains of duplicate 4KB blocks of some predefined
length. For each block to be written, it searches for all candidate block
chains containing the block and identifies the longest chain
for deduplication.  iDedup targets primary workloads rather than separate
backup images, so it has different design requirements.  Specifically, it does
not specifically optimize reads to latest data like ours.
%However, the search of block chains can become a bottleneck if there are many
%candidate chunk sequences.  
%Windows Server deploys a primary deduplication file system \cite{elshimi12}
%that uses a large block size for deduplication to reduce the indexing
%overhead, and applies block compression to maintain high storage efficiency. 

%{\bf Mitigating fragmentation.} Most deduplication storage systems suffer
%fragmentation, as pointed out in \cite{rhea08}. To mitigate fragmentation, 
%iDedup \cite{srinivasan12} deduplicates only a sequence of duplicate blocks
%whose length is larger than a predefined threshold, and amortizes the disk
%seeks in block reads.  A closely related work to ours is \cite{kaczmarczyk12},
%which proposes a backup system that mitigates fragmentation for more recent
%backups. Its idea is to selectively rewrite some deduplicated data into the
%disk, so as to reduce random I/Os in subsequent reads.  It uses a background
%process to remove old rewritten blocks.  Note that \cite{kaczmarczyk12} only
%focuses on mitigating fragmentation of different versions of the same backup
%stream, and does not consider deduplication across backup streams.  Also,
%\cite{kaczmarczyk12} only focuses on a single-disk system, and its scalability
%toward a disk-array system remains open. 

{\bf Version control systems.} Our work in essence provides similar
functionalities as in traditional version control systems.  Rdiff-backup
\cite{rdiffbackup} and Subversion \cite{svn} generate changes between adjacent
versions on a per-file basis.  In particular, Subversion improves restore
performance via a skip-list data structure.  Both studies do not
address global deduplication as in our work.  Git \cite{git} enables global
deduplication, but only in the whole-file level rather than the more
fine-grained block level. 
	
%\subsection{Challenges}
%\label{subsec:challenges}

% Most of the existing deduplication systems are designed exclusively for
% backing up a few specific servers. The assumptions about the system
% environment may not be applicable to a virtualization platform running
% multiple virtual machines in a number of VM hosts. In the following, we list
% out the challenges in designing a deduplication system for a virtualization
% platform with a large number of virtual machines.

% It is virtually impossible for having a raw disk read throughput for all
% backups, as deduplication causes data fragmentation. However not all backups
% are subject to reads in the future. For example backups older than one year
% should have a lower read opportunity than backups in a few days. Therefore by
% considering the system's overall performance, the read throughput of recent
% backups should match the match the raw disk throughput, while the patience on
% read aged backups can be greater.

\section{RevDedup}
\label{sec:design}

RevDedup is a deduplication system for backing up the disk states of
multiple VMs.  It builds on a client-server model similar to prior studies
\cite{clements09,guo11}. In RevDedup, a server stores deduplicated VM disk
images and the deduplication metadata, while multiple clients
run the active VMs operated by different users.  The server provides an
interface for each client to backup and restore specific VM images; the
clients take snapshots of VM disk images, compute fingerprints on
the snapshots, and upload the snapshots and fingerprints to the server.  

RevDedup considers a single snapshot created from the disk image of a VM as a
backup.  We call different snapshots that belong to the same VM to be {\em
versions}. RevDedup is designed to store and retrieve multiple versions of
different VMs in a virtualization environment. 

{\bf Goals.}
RevDedup aims to achieve several goals:
\begin{itemize}
\item {\em Storage efficiency:} It achieves high deduplication efficiency and
effectively reduces redundant storage of VM images. 
\item {\em Memory usage:} It uses limited memory usage for deduplication
indexing. 
\item {\em Backup}: It achieves high backup throughput of multiple backup
streams given the available resources in the system. 
\item {\em Restore:} It achieves near raw disk throughput in restoring the
latest versions of any VMs. 
\end{itemize}

%RevDedup aims to achieve high restore performance in deduplication storage,
%while maintaining high storage efficiency and high backup performance that have
%been extensively explored by prior studies.  
In this section, we elaborate how RevDedup achieves the above goals.  First,
to mitigate fragmentation on the read path, RevDedup applies 
{\em coarse-grained global deduplication} to amortize disk seeks over
large-size data units (see \S\ref{subsec:coarse}).  Also, to maintain high
deduplication efficiency, RevDedup further applies {\em fine-grained reverse
deduplication}, in which we maintain the data placement as sequential as
possible for the latest version, while removing any redundant data of the old
versions and referring it to the identical data of the latest version.  This
achieves high deduplication efficiency, and in the meantime mitigates
fragmentation and achieves high read performance for the latest version (see
\S\ref{subsec:fine}).  To improve scalability, our RevDedup implementation
offloads part of the deduplication workload from the server to multiple
clients and allows multiple clients to submit versions concurrently to the
server (see \S\ref{subsec:implementation}).   We also discuss how RevDedup
differs from conventional deduplication in backup/read performance (see
\S\ref{subsec:discussion}).

{\bf Assumptions.} We assume that RevDedup applies fixed-size chunking to
backup streams, i.e., we divide data into fixed-size units each identified by
a fingerprint, and determine if the unit can be deduplicated.  Fixed-size
chunking shows significant storage savings for VM images
\cite{jayaram11,jin09}, while having smaller chunking overhead than
variable-size chunking.  

We also assume that both RevDedup client and server processes run in user
space and are deployed in Linux.  The storage backend of the server is mounted
on a Linux native file system (e.g., Ext4 and XFS).  We leverage some
available functionalities of Linux in our design.  

Furthermore, RevDedup assumes that the stored data will never be deleted. The
issues of performing garbage collection on deleted versions are posed as
future work.

%Second, content-defined chunking (or variable-size chunking) creates
%unaligned chunk boundaries, and prevents our usage of some functionalities
%for reverse deduplication. We will further explain this in
%section~\ref{subsec:fine}.

\subsection{Coarse-Grained Global Deduplication}
\label{subsec:coarse}

The first approach that RevDedup uses to mitigate
fragmentation is to apply coarse-grained global deduplication to the pool of
the already stored VM snapshots (we discuss additional approaches in 
\S\ref{subsec:fine}).  By coarse-grained, we mean that RevDedup applies
deduplication to large fixed-size units called {\em segments}, each of which
has a size of several megabytes.  
%This is in contrast with the conventional deduplication approaches that
%usually use deduplication units on the order of kilobytes.  
By global, we mean that we apply deduplication to all versions and
eliminate duplicate segments that appear (i) in the same version, (ii) in
different versions of the same VM, or (iii) in different versions of different
VMs. Each segment is identified by a fingerprint that is generated from the
cryptographic hash of the segment contents. 

Our rationale of using large-size segments as our global deduplication units
is as follows. We expect that the content of a segment is sequentially written
to disk, and a disk seek occurs only if consecutive segments of a VM image are
not adjacently stored on disk due to deduplication.  With a large segment
size, the disk seek time of locating segments only forms a small portion of
the total time of reading all segment contents of a VM image.  Thus, we
effectively mitigate fragmentation by amortizing disk seeks over large-size
segments \cite{kruus10,srinivasan12}. 

Evaluations on our real-world dataset (see \S\ref{subsec:trace}) show that
using large-size segments for global deduplication can still achieve high
deduplication efficiency (with at least 80\% of space saving).
One possible reason is that files in a VM image are sequentially placed.
Changes of user files are likely aggregated in a small region, while the
operating system files remain intact.  Thus, the content differences of two
versions of the same VM are clustered in a small region of the VM image, 
and a substantial portion of segments will remain the same.  

Nevertheless, we point out that using large-size segments in deduplication
cannot maintain the same level of deduplication efficiency as in existing
fine-grained deduplication approaches. We address this in \S\ref{subsec:fine}.

\subsubsection{Indexing} 
\label{subsec:segindex}

The server holds a global deduplication index that keeps track of the
fingerprints and other metadata of all segments.  
%RevDedup does not exploit any temporal or spatial locality in the incoming
%backup streams to reduce the size of the index as in prior work (e.g.,
%\cite{}).  
By using large-size segments, the server can hold a small index that can be
fit into memory.  We justify this claim using a simple example.  Suppose that
the segment size is 8MB (a parameter used in our evaluation).  For each
petabyte of storage, we have to index 128 million entries.  Suppose that the
size of each entry is 32~bytes, which we believe suffice to store the
fingerprint (e.g., 20 bytes for SHA-1) and other metadata for each segment.
Then the index consumes a total size of 4GB, and can be fit into memory of
today's commodity hardware. 

\subsection{Fine-Grained Reverse Deduplication}
\label{subsec:fine}

In addition to segment-level deduplication, RevDedup also applies more
fine-grained deduplication on a sub-segment level to further eliminate
duplicates.  We define smaller fixed-size sub-segments called {\em blocks},
each of which has a size of several kilobytes. For example, the deduplication
block size can be set as the disk block size (e.g., 4KB) of native file
systems. Like segments, each block is identified by a fingerprint given by the
cryptographic hash of the block content.

RevDedup builds on a novel idea called reverse deduplication, which
mitigates fragmentation due to block-level deduplication in two ways. 
First, reverse deduplication is only local, meaning that it is only applied to
different versions of the same VM.  This avoids incurring disk seeks across
the versions of different VMs.  Second, and most notably, reverse
deduplication removes duplicate blocks of old versions and refers them to the
blocks of new versions.  This reduces the disk seeks of reading the latest
version of a VM and shifts the fragmentation overhead to older versions.

%This is in contrast with traditional deduplication approaches, in which new
%blocks are deduplicated with any old blocks that have already been stored.  

\subsubsection{Indexing}
\label{subsec:revindex}

Before discussing how reverse deduplication works, we first describe how we
perform indexing on block-level deduplication.  Each segment can be retrieved
from disk using the in-memory index (see \S\ref{subsec:segindex}).  It is
associated with a metadata file that is identified by the segment fingerprint.
The metadata file keeps the block fingerprints of all blocks associated
with the segment.  All metadata files are stored on disk.

For each version to be stored, RevDedup builds the index on the fly by loading
the metadata files of all segments into memory, and use this index for
block-level deduplication.  To quantify the memory usage, we consider the
following parameters used in our evaluation, such that the total size of a VM
image is 7.6GB, the block size is 4KB, and each block-level index entry is
32~bytes.  Since reverse deduplication operates by comparing similarities of
two versions of VM images (see below),  the total memory usage is up to
2$\times$7.6GB$\div$4KB$\times$32~bytes $=$ 121.6MB.  The actual memory usage
can be further reduced if we do not store the fingerprints of null
(zero-filled) blocks.  Also, if two segments are identical, their associated
blocks must also be identical and hence we do not need to load their block
fingerprints into memory.  Note that the index only temporarily resides in
memory and will be discarded after we finish deduplication for the version to
be stored.  

As long as the VM image size is on the order of tens of gigabytes, the memory
usage of our indexing approach is feasible with today's commodity hardware.
However, the memory usage increases proportionally with the VM image size.
One solution to reducing memory usage is to build the Bloom-filter-based
\cite{bloom70} index structure as in prior studies \cite{rhea08,zhu08}.  We
pose this issue as future work. 

%In addition, there are 7.6GB$\div$8MB $\approx$ 973 metadata files to be
%loaded.   We evaluate the index building time in \S\ref{sec:evaluation}. 

\subsubsection{Reverse Deduplication on Unique Segments}  
\label{subsec:reverseunique}

We first consider how reverse deduplication operates on different versions of
a single VM, assuming that all segments are unique and there is no global
deduplication across segments.  Note that different unique segments may still
share identical blocks.

\begin{figure}[t]
\centering
\includegraphics[width=0.4\linewidth]{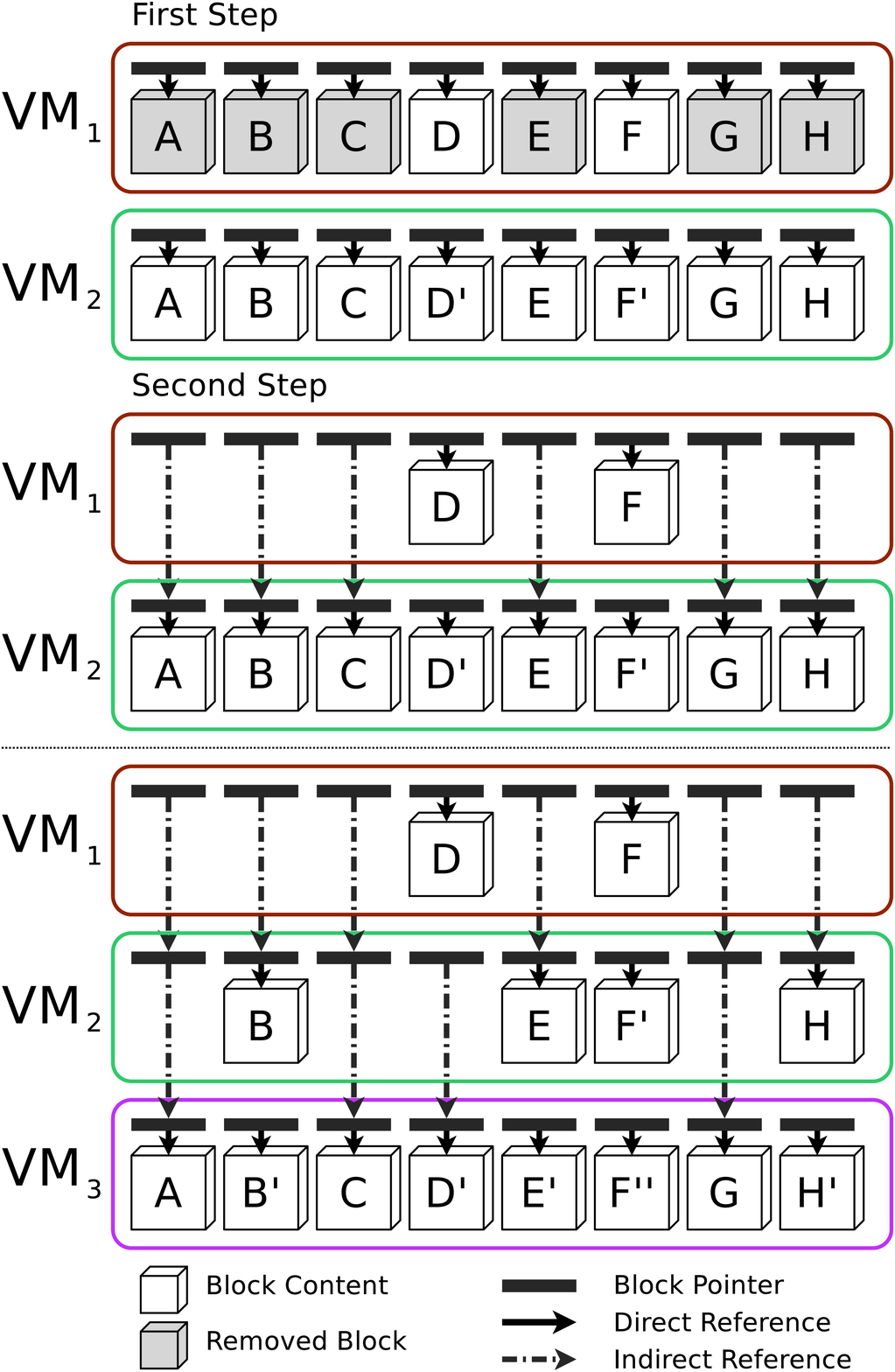}
\vspace{-6pt}
\caption{An example of reverse deduplication for multiple versions of
the same VM.}
\label{fig:reversededup}
\end{figure}

Figure~\ref{fig:reversededup} shows how reverse deduplication works, based on
the example shown in Figure~\ref{fig:eg_trad}.  Each version contains a number
of block pointers, each of which holds either a {\em direct reference}
to the physical block content of a segment, or an {\em indirect reference} to
a block pointer of a future version.  An indirect reference indicates
that the block can be accessed through some future version. In RevDedup, any
latest version of a VM must have all block pointers set to direct references. 

Suppose that the system has already stored a version VM$_1$, and now a new
version VM$_2$ of the same VM is submitted to the system.  We compare VM$_1$
and VM$_2$ by loading all their block fingerprints from disk. If a matched
block is found in both VM$_1$ and VM$_2$, we remove the respective block
of VM$_1$, and update that block with an indirect reference that refers to the
identical block of VM$_2$.  Now if we write another version VM$_3$ of the same
VM, we compare its blocks with those of VM$_2$, and remove any duplicate
blocks of VM$_2$ as above. Some blocks of VM$_1$ are now referred to those of
VM$_3$. To access those blocks of VM$_1$, we follow the references from
VM$_1$ to VM$_2$, and then from VM$_2$ to VM$_3$.

In general, when writing the $i$th version VM$_i$, we compare the block
fingerprints of VM$_i$ with those of the previous version VM$_{i-1}$.  We
remove any duplicate blocks of VM$_{i-1}$ and update the block pointers to
refer to the identical blocks of VM$_i$.  To simplify the deduplication
process, one key assumption we make is that we only compare with the most
recent version.  Hence, we may miss the deduplication with the redundant
blocks of earlier versions (i.e., VM$_{i-2}$, VM$_{i-3}$, $\cdots$, etc.).
Nevertheless, the analysis of our real-world dataset (see
\S\ref{subsec:trace}) indicates that such misses only contribute 0.6\% of
additional space usage.  Thus, RevDedup still achieves high deduplication
efficiency with this assumption.

When reading the $i$th version VM$_i$, we either follow the direct reference
to access the physical block, or a chain of indirect references to future
versions (i.e., VM$_{i+1}$, VM$_{i+2}$, $\cdots$, etc.) until a direct
reference is hit.  We point out that tracing the indirect references
incurs only small overhead in the read operation (see \S\ref{sec:evaluation}). 
	 
%the server loads the metadata files of all associated segments and identifies
%all blocks to read.  For each block to read, 

\subsubsection{Reverse Deduplication on Shared Segments}  
\label{subsec:reverse}

When segment-level global deduplication is in effect, we cannot directly
remove a block whose associated segment is shared by other versions
or within the same version.  RevDedup uses {\em reference counting} to
decide if a block can be safely removed. We associate each block with a
reference count, which indicates the number of direct references that
currently refer to the block among all versions of the same VM or different
VMs.  The block reference counts are kept inside the metadata files associated
with the segments. 
%so that the locality is preserved when RevDedup updates reference counts and
%removes duplicate blocks in reverse deduplication.

\begin{figure}
\centering
\begin{tabular}{c}
\includegraphics[width=0.6\linewidth]{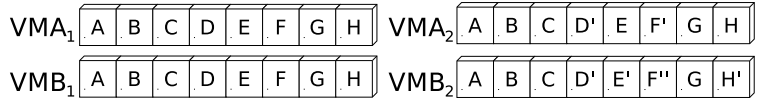}\\
\mbox{\small (a) VM versions}
\vspace{1em}\\
\includegraphics[width=0.4\linewidth]{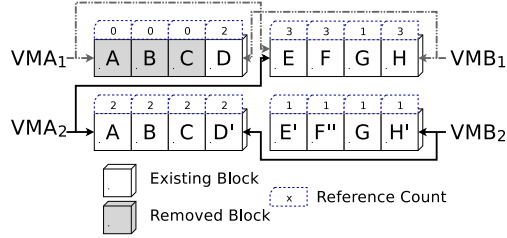}\\
\mbox{\small (b) Reference counts}
\end{tabular}
\vspace{-6pt}
\caption{An example that shows how reference counts are assigned when reverse
deduplication is applied to shared segments.}
\label{fig:refcnt}
\end{figure}

Figure~\ref{fig:refcnt} shows an example of how reference counts are used.
Suppose that two VMs, namely VMA and VMB, are stored.  Let the segment size 
be four blocks.  The first versions VMA$_1$ and VMB$_1$ have the same set of
blocks.  For the second versions, VMA$_2$ has new blocks {\sf D'} and 
{\sf F'}, while VMB$_2$ has new blocks {\sf D'}, {\sf E'}, {\sf F''}, and 
{\sf H'}.  We see that any blocks with zero reference counts (in the segment
{\sf ABCD}) can be safely removed. 

With reference counting, we now describe the complete reverse deduplication
design.  When a client writes the $i$th version VM$_i$ of a VM,
the server first applies global deduplication with the segments of other VMs.
For each segment of VM$_i$, if it is unique, then the reference counts of all
associated blocks are initialized to one; if the segment can be deduplicated
with some existing segment, then the reference counts of all associated blocks
of the existing segment are incremented by one.  Next, the server loads all
the block fingerprints of VM$_{i-1}$ (the previous version) and VM$_i$ into
memory.  It applies reverse deduplication and compares the block fingerprints
of VM$_{i-1}$ and VM$_i$. If a block of VM$_{i-1}$ can be deduplicated with
some block of VM$_i$, then the block of VM$_{i-1}$ will have its reference
count decremented by one and its direct reference updated to an indirect
reference that refers to the block of VM$_i$.  If the reference count reaches
zero, it implies that the block (of VM$_{i-1}$) is not pointed by any direct
references, but instead can be represented by the same block of future
versions.  It can thus be safely removed.

\subsubsection{Removal of Duplicate Blocks}
\label{subsec:removal}

RevDedup operates by removing duplicate blocks from segments.  We consider two
block removal approaches, namely block punching and segment compaction.
To this end, we propose a configurable mechanism that combines both approaches.

{\bf Block punching.} We leverage the hole-punching mechanism available
in Linux Ext4 and XFS file systems \cite{lwn}, where we can issue in user
space the system call {\tt fallocate(FALLOC\_FL\_PUNCH\_HOLE)} to a file
region.  Any file system block covered by the hole-punched region will be
deallocated and have its space released.  The respective block mappings of the
file will be updated in the file system.  

%while the remaining region inside a partially covered file system blocks will
%be zero-filled. 
%Since the unaligned spaces are not released by hole punching, RevDedup can
%only operate on fixed-size blocks for deduplication.

{\bf Segment compaction.}  Segment compaction is to compact a
segment that excludes the removed blocks.  It operates by copying all blocks
of a segment, except those that are to be removed, sequentially into a
different segment.  The original segment will be deleted and have its space
released, and the new segment is kept instead. 

%It also mitigates subsequent disk fragmentation, as the unique blocks now
%sits elsewhere in the file system.

%Yet the overhead of segment compaction will not be disastrous. In a
%worst-case situation, which RevDedup rebuilds a segment with all blocks
%unique, the file system has to read all the blocks inside that segment, and
%then writes them back elsewhere to the disk. Both operations are sequential,
%and thus in general the perfomance of segment compaction will be no less
%than half of the sequential disk throughput.

{\bf Threshold-based block removal.}  Block punching involves file
system metadata operations.  If the number of removed blocks is small, block
punching is expected to incur small overhead.  However, block punching has a
drawback of incurring disk fragmentation {\em (note that it is different
from the fragmentation problem in deduplication we discussed)}, as
non-contiguous free blocks will appear across disk.  This degrades write
performance when the amount of disk usage is close to its raw capacity.  On
the other hand, segment compaction mitigates disk fragmentation as it copies
all remaining blocks in sequence. However, if the number of removed blocks
is small, it has large I/O overhead since it reads and writes the actual data
content of the non-removed blocks. 
% (although both reading and writing of segments are sequential operations).  
%Nevertheless, we argue that the overhead remains small.  In the worst case,
%the rebuilt segment contains only one removed block, and there will be %one
%read (of the original segment) and one write (of the new segment).  Both
%operations are sequential and remain fast.  

%For a segment under reverse deduplication, if it contains few blocks to be
%removed, then block punching should yield better performance as RevDedup needs
%to read and write a large number of blocks in segment compaction.  Also, the
%adverse impact of disk fragmentation is small.  On the other hand, if the
%segment contains many blocks to be removed, disk fragmentation becomes severe,
%while segment compaction has limited overhead as only few blocks are copied.
Therefore, we propose a threshold-based block removal mechanism,
which uses a pre-defined threshold (called the {\em rebuild threshold}) to
determine how to rebuild a segment excluding removed blocks.  If the fraction
of blocks to be removed from a segment is smaller than the rebuild threshold,
then block punching will be used; otherwise, segment compaction will be used.
The rebuild threshold is configured to trade between disk fragmentation and
segment copying time.  We evaluate the impact of the rebuild threshold in
\S\ref{sec:evaluation}. 

Note that after we remove some blocks from a segment, no more blocks will be
further removed from the same segment.  In other words, we only apply block
removal (via either block punching or segment compaction) to a segment 
{\em at most once} only.  If a segment contains blocks with zero reference
counts, it implies that no latest versions refer to the segment.  For example,
from Figure~\ref{fig:refcnt}, when we remove blocks from the segment 
{\sf ABCD}, only the old versions VMA$_1$ and VMB$_1$ refer to it.  When we
upload future versions of VMA (or VMB), the segment will no longer be
compared, while only the segments referenced by the latest version VMA$_2$ (or
VMB$_2$) will be considered. 

%We evaluate the performance impact of different threshold values in
%\S\ref{sec:evaluation}.

%In practice block punching and segment compaction only contributes a small
%fraction of deduplication overhead. From our dataset (see
%Section~\ref{sec:evaluation}), we observe decent deduplication
%efficiency when we first apply coarse-grained global deduplication at the
%segment level (see Section~\ref{subsec:coarse}). This implies that we can
%discard a large number of duplicate segments on the write path and no longer
%need to perform reverse deduplication on them.  Thus, we expect that the
%overhead is small and has limited impact on the deduplication process.

\subsection{Implementation}
\label{subsec:implementation}

\begin{figure}[t]
\centering
\includegraphics[width=0.6\linewidth]{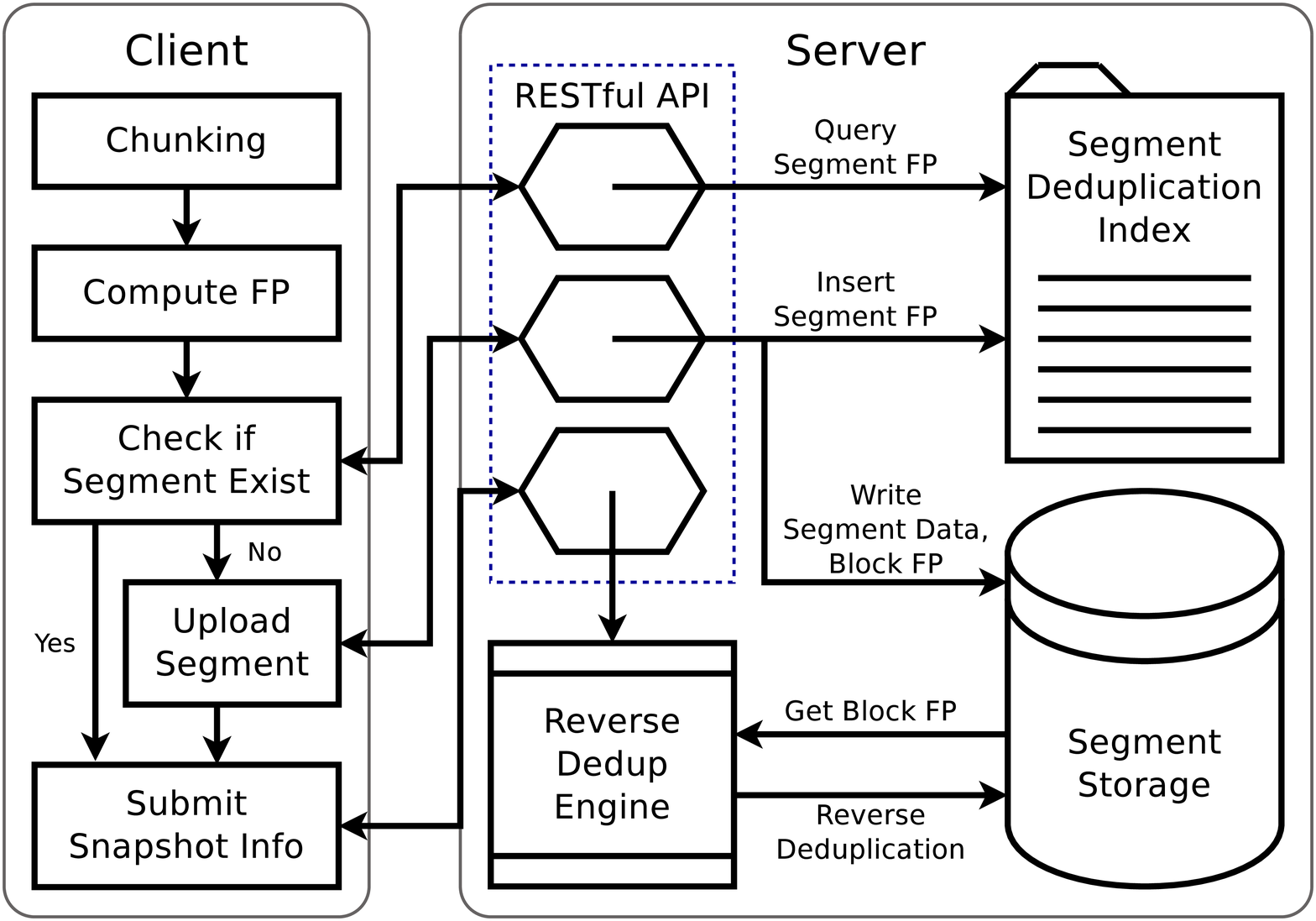}
\vspace{-6pt}
\caption{RevDedup's client-server model.}
\label{fig:model}
\end{figure}

Our RevDedup implementation builds on the client-server model as shown in 
Figure~\ref{fig:model}.  
RevDedup uses client-side deduplication to reduce the client-server
communication overhead.  When a client is about to submit a version of
a VM to the server, it first divides the VM image snapshot into different
segments and computes both segment-level and block-level fingerprints for the
version.  Next, the client queries the server, using the segment fingerprints,
whether the segments are already stored in the server.  If any segment has
already been stored, then the client discards the upload of that segment.
The client then uploads the unique segments to the server (e.g., via RESTful
APIs). It also sends the metadata information, including all segment and block 
fingerprints for the whole VM image and the information of the version (e.g.,
the VM that it belongs, the size of the image, etc.). Note that we offload the
server by having the clients be responsible for both segment and block
fingerprint computations.  This avoids overloading the server when it is
connected by too many clients.

Upon receiving the unique segments and metadata information of a version, the
server writes them to disk and links the version with the existing segments
that are already stored.  The server performs reverse deduplication as
described in \S\ref{subsec:fine}, including: loading metadata files and
building the block fingerprint index on the fly, searching for duplicates and
updating direct/indirect references, and removing duplicate blocks from
segments via block punching or segment compaction. 

%complete. Therefore the server can handle segment upload request of a version
%and performs reverse deduplication on another version concurrently.

%The server first loads all the block fingerprints in the segments owned by the
%version to be stored and builds the block fingerprint index in memory. Next,
%the server loads all the block fingerprints of the previous version, and
%searches the duplicate blocks in the previous version. If a block is found
%duplicate, RevDedup updates its direct reference to be indirect reference. At
%last RevDedup removes the duplicate blocks in the segments of the previous
%version.

Our RevDedup prototype is implemented in C in Linux. We use SHA-1 for both
segment and block fingerprint computations.  The RevDedup server mounts its
data storage backend on a native file system.  RevDedup requires that the file
system support hole-punching, and here we use Ext4 for Linux.  In the
following, we address several implementation details.

{\bf Mitigating interference.}
Since a client may perform fingerprint computations for a running VM,
minimizing the interference to the running VM is necessary.  Here, we can
leverage the snapshot feature that is available in today's mainstream
virtualization platforms.  The client can directly operate on the mirror
snapshot in the background and destroy the snapshot afterwards.

%The server uses GNU {\tt libmicrohttpd} 
%\cite{libmicrohttpd} as the HTTP server frontend, while each client uses
%{\tt libcurl} \cite{libcurl} to generate HTTP requests. 
{\bf Communication.}
The client-server communication of RevDedup is based on RESTful APIs, which
are HTTP-compliant.  A client can retrieve a VM image by issuing a correct
HTTP request.  The server can process multiple requests from different
simultaneously.  

{\bf Multi-threading.} RevDedup exploits multi-threading to achieve high
read/write performance.  
In writes, the server uses multiple threads to receive segment uploaded by the
clients and to perform reverse deduplication.  In reads,
the server uses dedicated threads to pre-declare the segment reads in kernel
by using the POSIX function {\tt posix\_fadvise(POSIX\_FADV\_WILLNEED)}. With
read pre-declaration, the kernel can make effective pre-fetching of segments
to improve read performance.   

In addition, the server uses a separate thread to trace the chains of indirect
references of blocks when old versions are read.  Once the direct reference is
found, the thread sends the block address to another thread for reading the
block content.  Both threads run concurrently.  This reduces the overhead
of tracing long indirect reference chains. 
	 
%We compare the time required to rebuild the reference chains and to read the
%versions in \S\ref{sec:evaluation}.

{\bf Handling of null blocks.} In practice, VM images contain a large number
of null (or zero-filled) blocks \cite{jin09}.  In RevDedup, the server skips
the disk writes of any null blocks appearing in the segments submitted by a
client.  When a null block is to be read, the server generates null data on
the fly instead of reading it from disk.

\subsection{Discussion}
\label{subsec:discussion}

Conventional (inline) deduplication (e.g., in \cite{quinlan02,rhea08,zhu08})
typically applies global deduplication to small-size data units and removes
duplicates from new data.  It is
equivalent to setting a small segment size for global deduplication and
disabling reverse deduplication in RevDedup.  Conventional deduplication
generally achieves higher backup throughput than RevDedup, since it can
discard more duplicate segments on the client side with more
fine-grained global deduplication, while RevDedup removes some duplicates 
on the server side via reverse deduplication.  Nevertheless,
RevDedup can still achieve high backup throughput if the client side discards
enough duplicate segments with our coarse-grained global deduplication.  On the
other hand, conventional deduplication sees decaying read performance for
newer data due to fragmentation.  We compare both approaches in
\S\ref{sec:evaluation}.

\section{Evaluation}
\label{sec:evaluation}

We conduct testbed experiments on RevDedup using different workloads,
including unique data and two VM datasets.  Our results are summarized as
follows:
\begin{itemize}
\item 
RevDedup maintains high baseline throughput in unique data
(\S\ref{subsec:unique}).
\item
We show via real-world VM traces that RevDedup achieves: (i) high
deduplication efficiency compared to existing deduplication approaches, (ii)
high backup throughput given the available resources in the system, and (iii)
high read throughput for restoring the latest versions of any VMs
(\S\ref{subsec:trace}).
\item
We show via a VM backup with a long version chain that: (i) RevDedup incurs
small metadata overhead in the reverse deduplication process and can be
configured between block punching and segment compaction during backup
operations, and (ii) RevDedup incurs small overhead in tracing indirect
references for earlier versions (\S\ref{subsec:synthetic}). 
\end{itemize}

%Our goal is to evaluate RevDedup in both micro and macro levels. 
%We first test RevDedup using synthetic workloads.  We further evaluate the
%performance of RevDedup using real-life VM image datasets.
Our experiments are conducted on a machine with a 3.4GHz Intel Xeon
E3-1240v2 quad-core, eight-threaded processor, 32GB RAM, and a RAID-0 disk
array with eight ST1000DM003 7200RPM 1TB SATA disks \cite{seagate}.  Since our
testbed has only 8TB of raw storage, the actual memory usage of RevDedup in
our experiments is much less than 32GB based on our calculations in
\S\ref{subsec:segindex} and \S\ref{subsec:revindex}.  Also, we point out that 
RAID-0 is not recommended for fault-tolerant systems as it stripes data
without parities, but we choose it in our experiments to maximize the disk
array throughput for our stress tests.  The machine runs Ubuntu 12.04 with
Linux kernel 3.2.0. 

We create eight client processes and one server process, all of which are
executed on the same machine and are connected by the Linux loopback
interface, so as to eliminate the network bottleneck for our high-performance
benchmarking.  The client processes submit VM data to the server concurrently. 

We consider four segment sizes for global deduplication: 4MB, 8MB, 16MB, and
32MB. We fix the block size at 4KB for reverse deduplication to match the
file system block size.  We set the default rebuild threshold at 20\% for our
block removal mechanism.
%We also evaluate other thresholds in \S\ref{subsec:synthetic}. 
%For evaluating the threshold, we focus on 5 difqferent values, which ranges
%from 0\% to 40\% with 10\% level increment. In other experiments, we fix the
%threshold value to be 20\%.

We focus on examining the read/write performance of RevDedup.  We exclude
fingerprint calculations from our experiments.  In real deployment, the
clients can generate VM snapshots and compute fingerprints offline before
connecting to the server.  The fingerprint computations should not affect the
actual read/write performance of RevDedup.  In our experiments, we pre-compute
all segment and block fingerprints before benchmarking.  
%When we benchmark the read performance, we clear the kernel page cache before
%each read and discard any read data by redirecting it to the Linux null
%device, so that the measured read performance is due to disk accesses.
Our throughput results are averaged over five runs.

\subsection{Evaluation with Unique Data}
\label{subsec:unique}

%Before we evaluate RevDedup using snapshots from a live VM instance, 
%the base performance of RevDedup with unique data. We
We measure the baseline performance of RevDedup using unique data.  The server
initially contains no data.  The client processes submit 128GB of
unique data (i.e., all blocks are globally unique) to the server.  Then a
client process retrieves the data using the Linux command {\tt wget}.  We also
measure the raw disk throughput by reading/writing data directly via the
native file system of our testbed.  

Table~\ref{tbl:unique} shows the results.  The write throughput of RevDedup is
13-19\% less than the raw write throughput, mainly because RevDedup needs to
handle segment metadata including fingerprints and reference counts.  When the
segment size is larger, fewer segments are involved and RevDedup has higher
unique write throughput.  On the other hand, the read throughput of RevDedup
is very close to the raw read throughput. 

\begin{table}[!h]
\centering
\begin{tabular}{|c|c|c|c|c|c|}
\hline
(GB/s) 	& Raw & 4MB & 8MB & 16MB & 32MB \\
\hline
Write	& 1.37 & 1.11 & 1.18 & 1.17 & 1.20  \\
Read 	& 1.27 & 1.26 & 1.27 & 1.27 & 1.25  \\
\hline
\end{tabular}
\caption{Throughput of RevDedup for unique data.}
\label{tbl:unique}
\end{table}

\subsection{Evaluation with Real-World VM Usage}
\label{subsec:trace}

We evaluate RevDedup for backing up the disk states of multiple VMs based on
real-world workloads.  

\subsubsection{Dataset}

We collected a real-world dataset from the snapshots of VMs used by university
students in a computer science programming course. We prepared a master image
of size 7.6GB with 32-bit Ubuntu 10.04 installed. We then cloned 160 VMs, and
assigned one to each student to work on three programming assignments in a
semester.  We generated 12 weekly versions for each VM, and computed a
cryptographic hash for every 4KB block in each version.  If no deduplication
is applied, the total size of all versions over the 12-week span is 14.3TB.
If we exclude null blocks, there is 6.67TB of data.
	
\begin{figure}[h]
\centering
\includegraphics[width=0.5\linewidth]{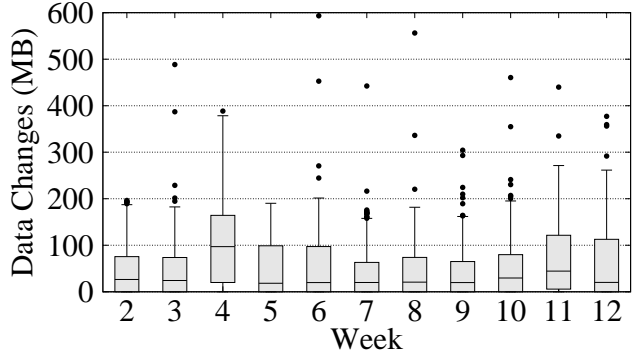}
\vspace{-1em}
\caption{Boxplots of data changes (we only plot data within 600MB).}
\label{fig:boxplots}
\end{figure}

We first analyze the dataset to develop ground truths.
Figure~\ref{fig:boxplots} shows the boxplots for the distributions of changes
of each weekly version (from Week~2 to Week~12) with respect to the
version of the same VM in the previous week.  Each boxplot shows the minimum,
lower quartile, medium, upper quartile, and maximum of all 160 versions each
week.  We note that most VMs have less than 100MB of changes per week.  In
Week~4, there is a spike of data changes due to an assignment deadline. Our
dataset also contains outliers that generate significant data changes.  For
example, in Week~12, a student generates 6GB of new data (not shown in the
figure).

\begin{figure*}[t]
\begin{minipage}{0.325\linewidth}
	\centering
	\includegraphics[width=\linewidth]{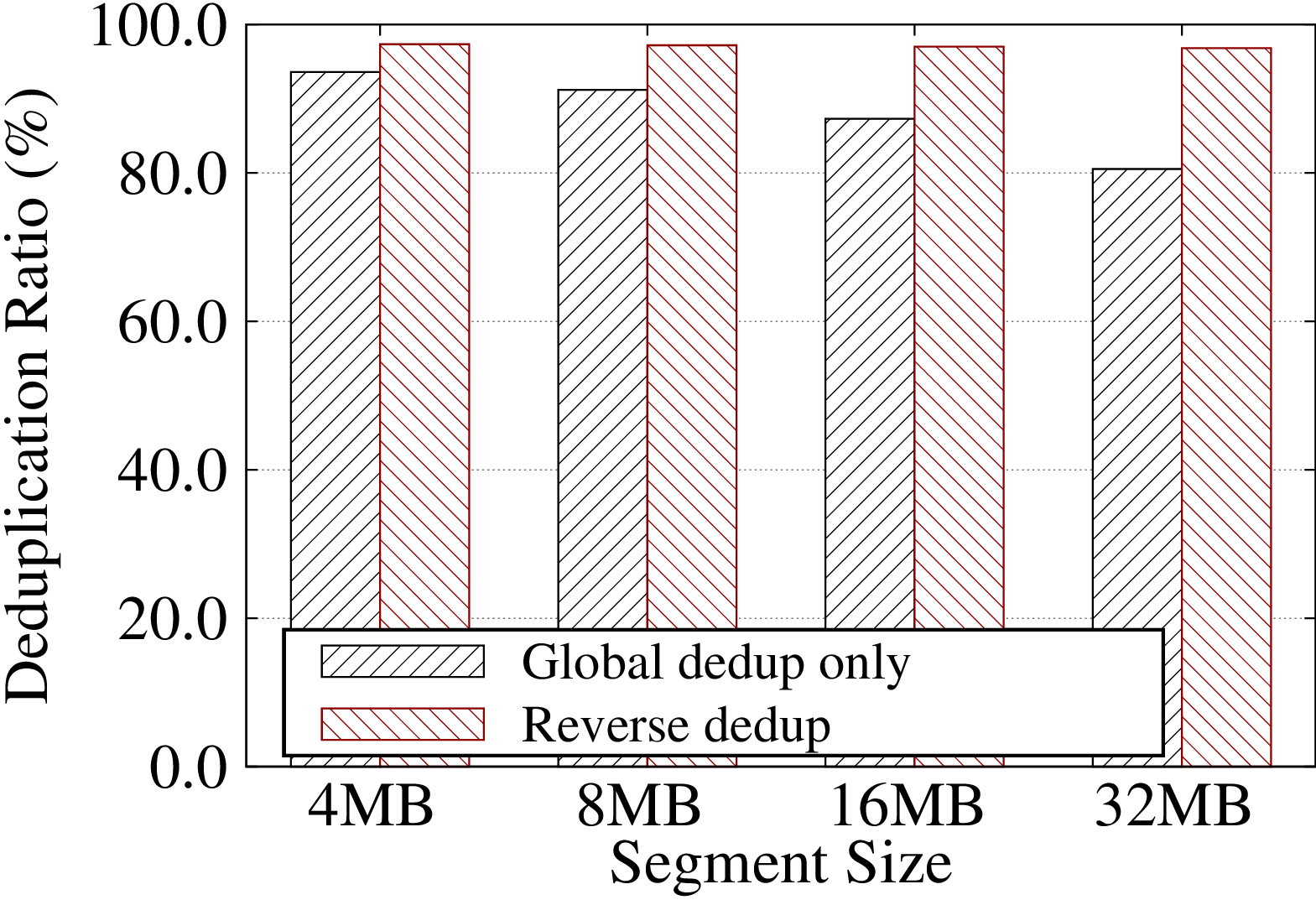}
	\parbox[t]{0.95\linewidth}{\small (a) Dedup ratios of reverse dedup and
		conventional dedup}
\end{minipage}%
\hspace{0.005\linewidth}
\begin{minipage}{0.325\linewidth}
	\centering
	\includegraphics[width=\linewidth]{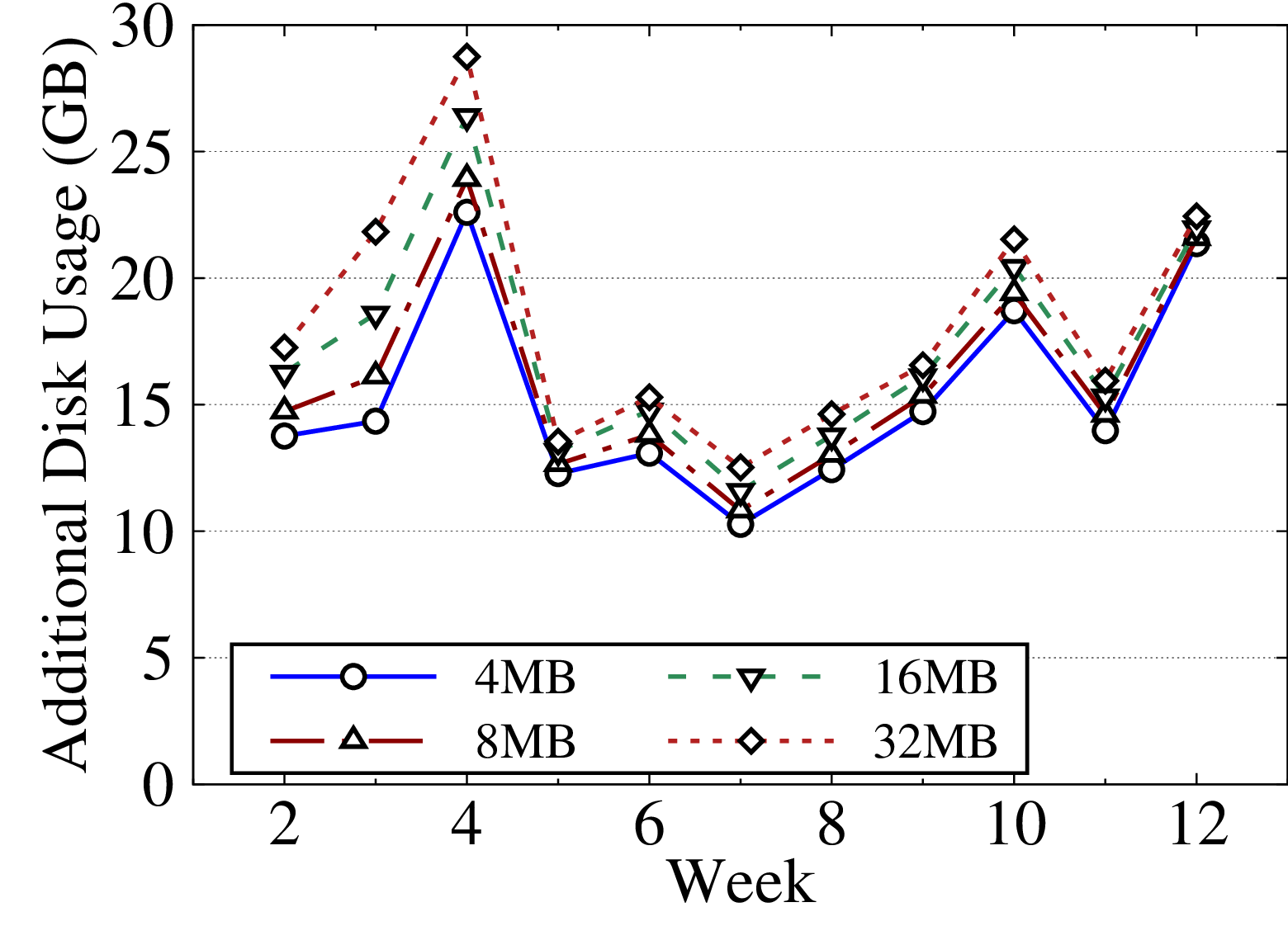}
	\parbox[t]{0.95\linewidth}{\small (b) Additional disk usage for storing versions using RevDedup}
\end{minipage}%
\hspace{0.005\linewidth}
\begin{minipage}{0.325\linewidth}
	\centering
	\includegraphics[width=\linewidth]{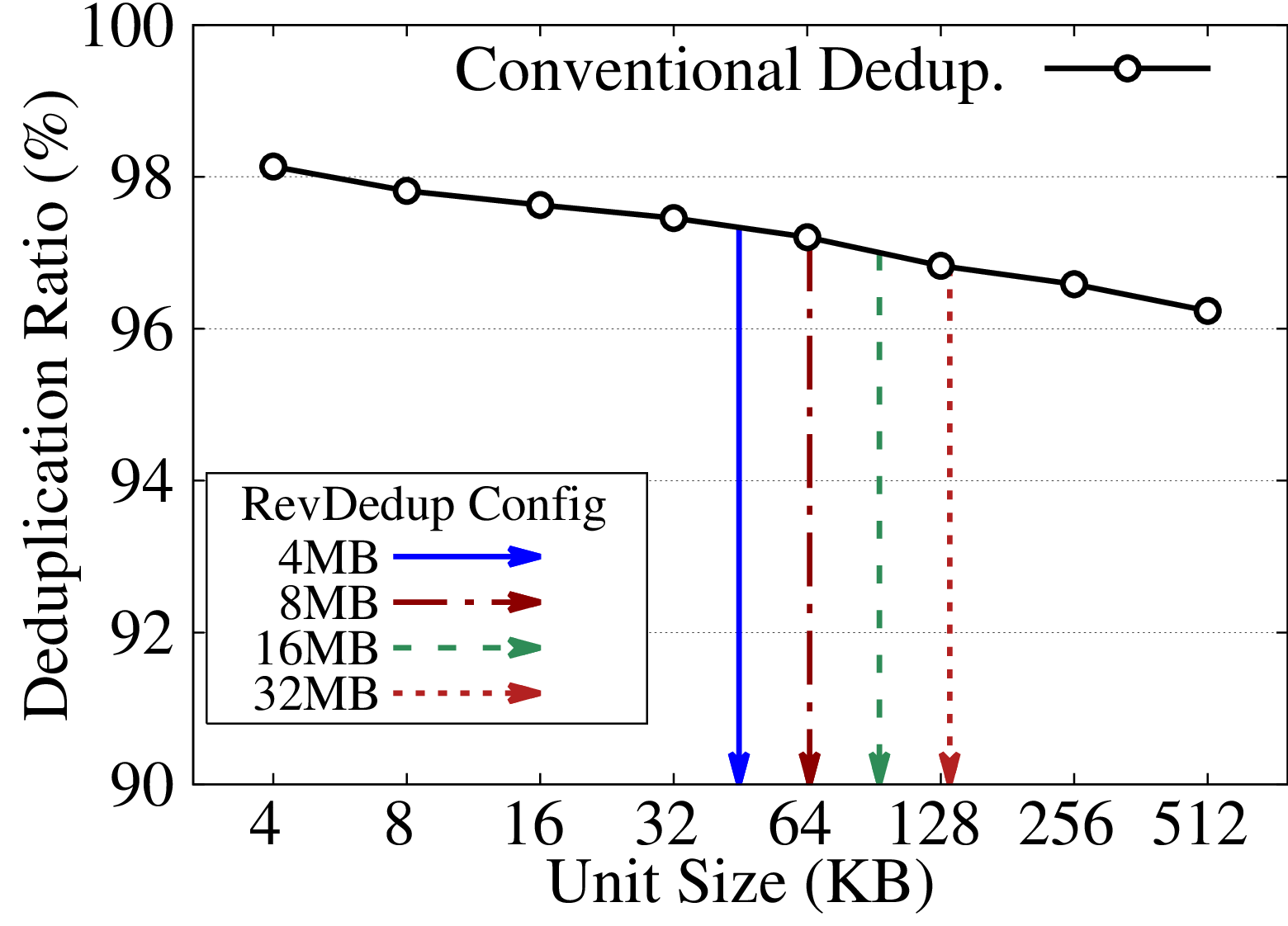}
	\parbox[t]{0.95\linewidth}{\small (c) Dedup ratios of RevDedup and
		conventional dedup}
\end{minipage}
\caption{Deduplication efficiency of RevDedup in real-world workloads.}
\label{fig:efficiency}
\end{figure*}

%For privacy reasons, we keep only the block hashes and discard the content of
%the version.  When benchmarking the read and write performance, we reconstruct
%each block by placing its block hash at the front of a null (zero-filled) 4KB
%block.  Any two identical (resp. different) block hashes will give identical
%(resp. different) blocks.  This preserves the characteristics of content
%similarities in the granularity of 4KB.

\subsubsection{Storage Efficiency}
\label{subsec:efficiency}

We evaluate the storage efficiency of RevDedup when storing the 12 weekly
version sets.  We measure the actual disk usage including both data and
metadata.

{\bf Effectiveness of reverse deduplication.} We examine the storage savings
of RevDedup using reverse deduplication (with 4KB block size).  We define the
{\em deduplication ratio} as the percentage of space saved with deduplication
to the total size of all VM images (excluding null blocks) without
deduplication.  A higher deduplication ratio means higher storage efficiency.
Here, we compare two variants of RevDedup: (i) only coarse-grained global
deduplication is used, and (ii) both global and reverse deduplication
approaches are used.  Figure~\ref{fig:efficiency}(a) shows the deduplication
ratios.  Coarse-grained global deduplication itself achieves space saving of
80.5-93.6\%, while reverse deduplication further removes duplicates and
increases the saving to 96.8-97.3\%.  This saving is comparable to existing
deduplication systems (see below).  In the following experiments, we
enable reverse deduplication in RevDedup. 

{\bf Additional space usage per week.}  We now provide a more detailed
analysis of the space usage of RevDedup for each weekly version set.
Figure~\ref{fig:efficiency}(b) shows the additional disk space for storing
each weekly version set using RevDedup since Week~2.  We see that the trend
follows that of the change distributions shown in Figure~\ref{fig:boxplots}
(e.g., large space usage in Week~4).  Note that RevDedup introduces
more additional space in Week~12 than in Week~11, although both weeks have
similar change distributions (see Figure~\ref{fig:boxplots}). The reason is
that Week~12 has outliers that make significant changes.  A key observation is
that the additional space usage only increases marginally when the segment
size increases from 4MB to 32MB. 
%Hence, RevDedup still maintains high storage efficiency even when using a
%large segment size. 
%It confirms our conjecture that two versions of the same VM have content
%differences clustered in a small region (see \S\ref{subsec:coarse}), and
%hence a large segment size still maintains high storage efficiency.

{\bf Comparisons with conventional deduplication.}  We now compare the storage
efficiency of RevDedup with conventional deduplication that operates on data
units of small size (e.g., few kilobytes).
Figure~\ref{fig:efficiency}(c) plots the deduplication ratios.  The various
configurations of RevDedup have similar storage efficiency to the conventional
approaches with data unit size ranging from 32KB to 128KB.  The results
indicate that RevDedup can achieve comparable storage efficiency to some
state-of-the-art deduplication file systems.  For example, ZFS \cite{zfs} and
Opendedup SDFS \cite{opendedup} operate on fixed-size units with default size
128KB. 

% We further examine the storage efficiency improvements via fine-grained
% reverse deduplication.  We define the {\em deduplication factor} as the 
% ratio of the original space usage of all VM images (excluding null blocks)
% without deduplication to the resulting space usage after deduplication is
% applied. A higher deduplication factor means higher storage efficiency. 
% 
% Figure~\ref{fig:dedupfactor}(a) shows the deduplication factors with and
% without reverse deduplication.  Coarse-grained global deduplication itself
% achieves a deduplication factor of 5:1 to 12:1, while reverse deduplication
% further improves the storage efficiency to a deduplication factor of over
% 30:1.  It confirms that reverse deduplication is effective to further remove
% duplicates between VM backup versions. 

\begin{figure*}[t]
\centering
\begin{minipage}{0.32\linewidth}
\centering
\includegraphics[width=\linewidth]{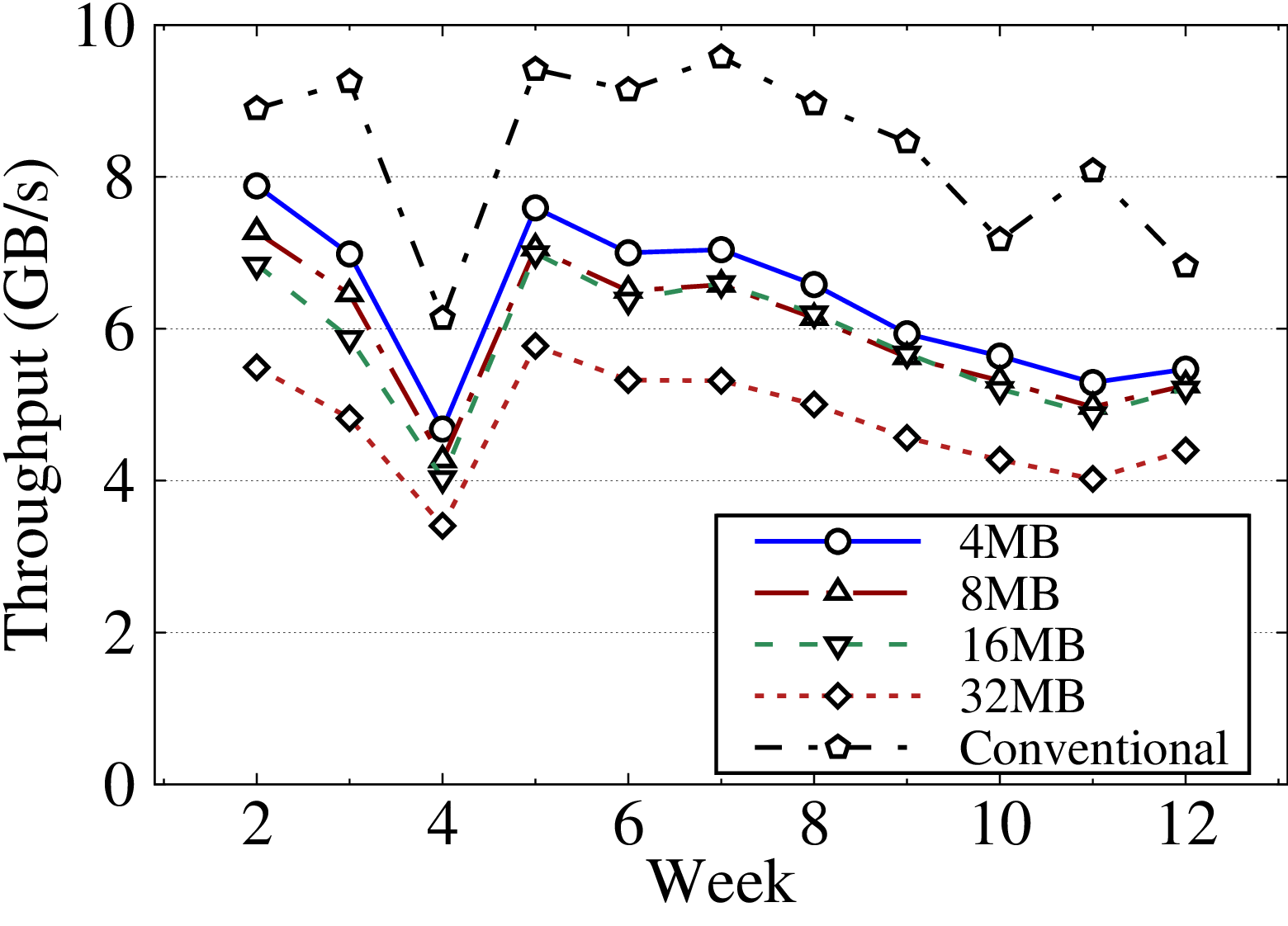}
\mbox{\small (a) Storing versions}
\end{minipage}
\begin{minipage}{0.32\linewidth}
\centering
\includegraphics[width=\linewidth]{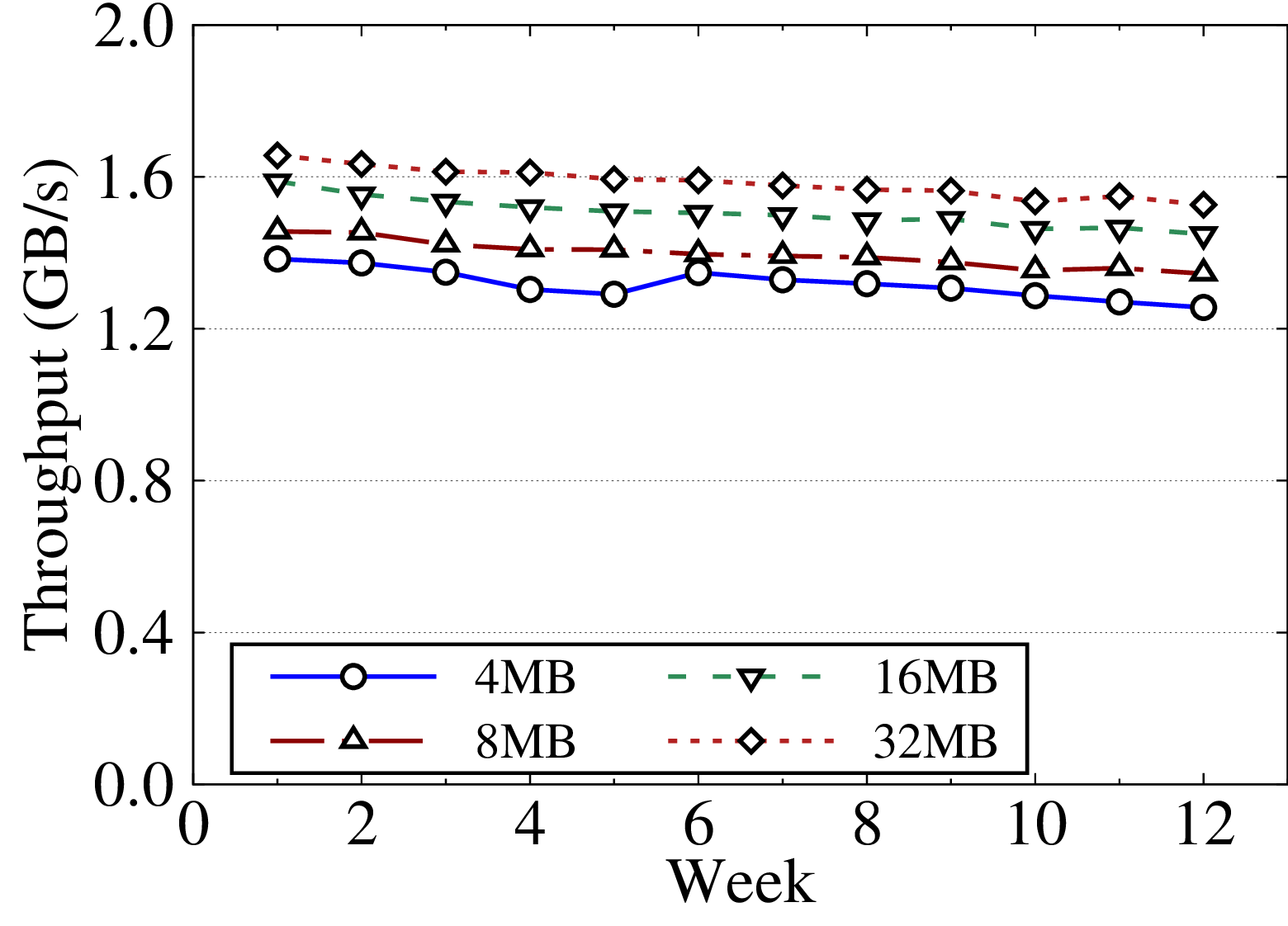}
\mbox{\small (b) Reading latest versions}
\end{minipage}
\begin{minipage}{0.32\linewidth}
\centering
\includegraphics[width=\linewidth]{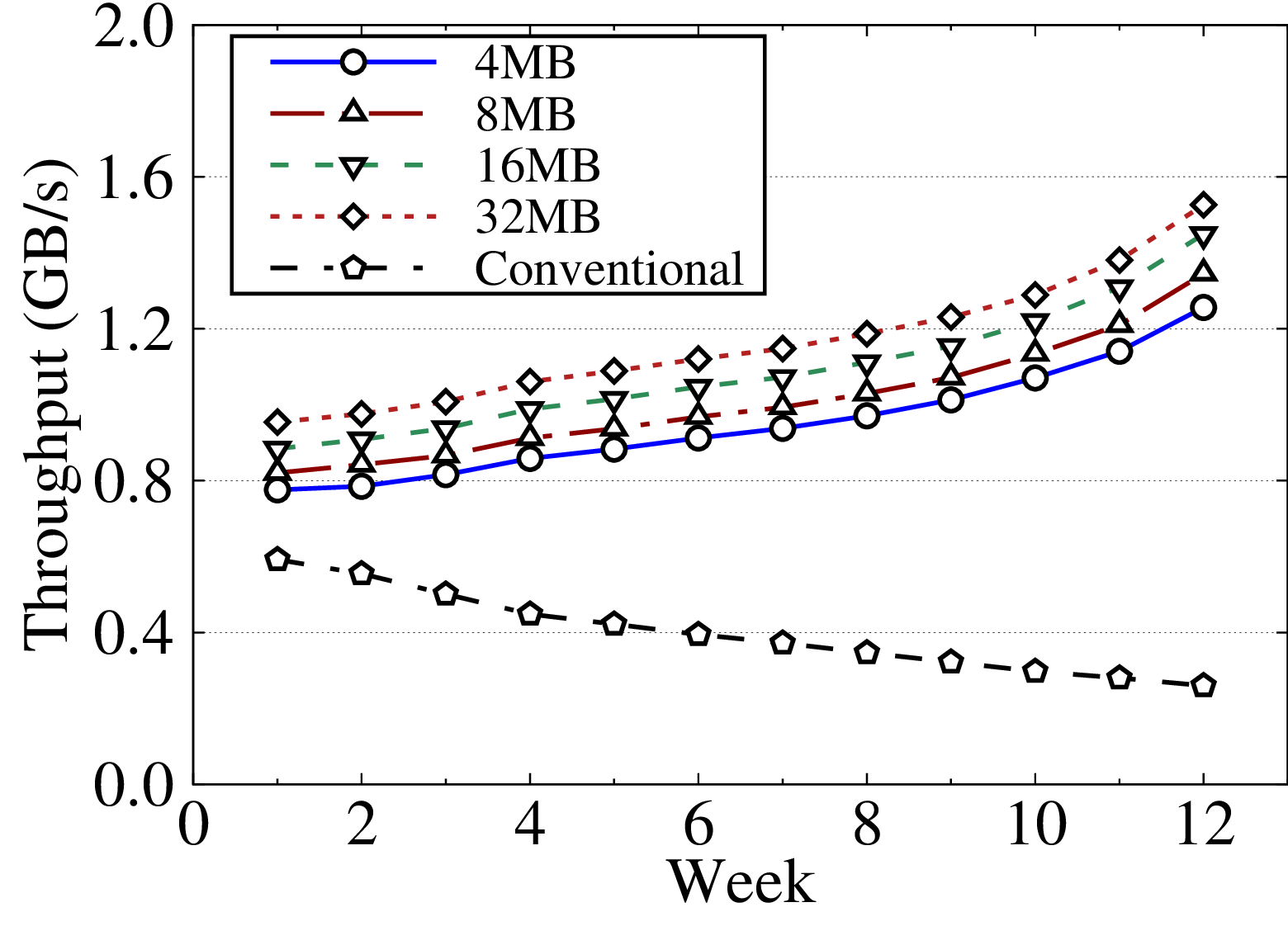}
\mbox{\small (c) Reading earlier versions}
\end{minipage}
\caption{Performance of RevDedup when storing/reading 12 weekly version sets.}
\label{fig:traceops}
\end{figure*}

\subsubsection{Throughput}

We compare the backup and read throughput of RevDedup for different segment
sizes and that of conventional deduplication.  To evaluate the latter, we
configure RevDedup to use a 128KB segment size for global deduplication and
disable reverse deduplication.  As shown in \S\ref{subsec:efficiency}, the
128KB segment size has comparable deduplication efficiency to RevDedup and is
the default setting in some state-of-the-art deduplication file systems
\cite{zfs,opendedup}.  We only modify the chunking configurations of RevDedup
to resemble a conventional deduplication approach, while retaining other
implementation features described in \S\ref{subsec:implementation} (e.g.,
multi-threading and handling of null blocks).  This enables us to compare
RevDedup and conventional deduplication under fair conditions. 

{\bf Backup throughput.}  We evaluate the backup throughput of storing the 12
version sets.  The server has no data initially.  Then we submit the 12
version sets in the order of their creation dates.  We measure the time of the
whole submission operation, starting from when the clients submit all unique
segments until the server writes them to disk and performs reverse
deduplication (for RevDedup).  We call {\tt sync()} at the end of each write
to flush all data to disk.  Here, we plot the results starting from Week~2, in
which RevDedup begins to apply reverse deduplication to the version sets being
stored.

Figure~\ref{fig:traceops}(a) shows the throughput of RevDedup and
conventional deduplication in the backup of each weekly version set.
As discussed in \S\ref{subsec:discussion}, conventional
deduplication has higher backup throughput than RevDedup, for example, by an
average of 30\% compared to RevDedup with segment size 4MB.   Nevertheless,
RevDedup
still achieves high backup throughput in the range around 4-7GB/s.  Its backup
throughput is higher than the raw write throughput by around 3-5$\times$ as it
can discard the writes of a large number of duplicate segments on the client
side (as shown from the storage saving of our coarse-grained global
deduplication in Figure~\ref{fig:efficiency}(a)).  A smaller segment
size implies higher throughput since more duplicates are discarded on the
write path.  Note that there is a throughput drop in Week~4 due to significant
modifications made to the VMs (see Figure~\ref{fig:boxplots}). 

%The results of Figure~\ref{fig:traceops}(a) provide some insights into the
%impact of the segment size.  When the segment size is smaller, more
%duplicates are discarded on the write path and this leads to smaller write
%and reverse deduplication times,  but more segments are managed and this
%increases the metadata processing time.  Overall, we see that a smaller
%segment size yields higher backup throughput. 

{\bf Read throughput.}  We evaluate the read throughput of RevDedup in two
parts.  We submit each of the 12 version sets in the order of their creation
dates. After submitting a version set, we immediately retrieve all versions of
that set (call it ``reading latest versions'').  After submitting all
versions, we measure the time of reading each version set (call it ``reading
earlier versions'').  Note that conventional deduplication has the same read
throughput in both parts as it does not modify the earlier versions after they
are stored (as opposed to RevDedup).  Before each measurement, we flush the
file system cache using the command 
``{\tt echo 3 > /proc/sys/vm/drop\_caches}''.

Figure~\ref{fig:traceops}(b) shows the throughput of RevDedup in reading the
latest versions.  The overall read throughput of RevDedup is 1.2-1.7GB/s.
The latest versions are only subject to segment-level fragmentation due to
coarse-grained global deduplication.
For the version sets of later weeks, the read throughput drops as the
degree of segment-level fragmentation increases, but the drop remains small.
A larger segment size also gives higher read throughput as the fragmentation
is better amortized.  We point out that the read throughput of RevDedup is
higher than the raw read throughput.  The reason is that RevDedup generates
null blocks of VM images on the fly rather than reading them from disk (see
\S\ref{subsec:implementation}).  We expect that as a VM ages, more null blocks
will be filled with content and eventually RevDedup will see read throughput
drop below the raw read throughput. 
%causing a 10\% drop for segment configurations. We believe that %the read
%throughput of RevDedup is below the raw-disk throughput.
Therefore, {\em we do not claim that RevDedup reads faster than raw read}.
Nevertheless, the throughput drop is expected to be mild as we use a large
segment size to amortize disk seeks.

%  Thus, for the later weeks, the read throughput drops as the
% degree of segment-level fragmentation increases. qHowever, the drop remains
% small, especially for the 16MB and 32MB segment sizes.
% Their drops from Week~1 to Week~12 are less than 10\% and 8\%, respectively.

Figure~\ref{fig:traceops}(c) shows the throughput of RevDedup in reading
earlier versions after storing all versions.  We also include the results of 
conventional deduplication here.  We observe that RevDedup confirms our design
goal, as the read throughput decreases with earlier versions being read.  For
example, the read throughput for Week~1 is up to 40\% less than that for
Week~12.  The figure also shows the fragmentation problem in conventional
deduplication.  For Week~1, the read throughput can only achieve 606MB/s (at
least 25\% less than RevDedup), mainly due to fragmentation introduced in
global deduplication with the small segment size at 128KB.  The read
throughput decreases further for later weeks. It drops to 266MB/s for Week~12,
which is only around 20\% of the raw read throughput (see
Table~\ref{tbl:unique}).  This shows that fragmentation becomes more severe
for newer versions.  

\subsection{Evaluation with a Long-Chained VM}
\label{subsec:synthetic}

We now evaluate the metadata overheads of backup/read operations in RevDedup
when the number of versions of a VM backup grows.  

%The dataset provides implications on how RevDedup performs when reading a
%very old version.  We examine RevDedup in deduplication overhead and segment
%rebuilding.  %In particular, we show how segment rebuilding can be configured
%with different %thresholds to trade between reverse deduplication time and
%disk %fragmentation. 

\subsubsection{Dataset}

The dataset was also used in the prior work \cite{tang12}.  
We consider a Fedora 14 VM configured with 5GB disk space. The VM ran a cron
job ``{\tt yum -y update}'' to download and install updates daily.  The
updates modified the VM system files accordingly.  The VM also ran other
background maintenance jobs that may change the disk state, but did not
generate any user data.  We collected 96 daily versions of the VM.  We 
find that the VM has around 50-100MB changes of data per day. 

\subsubsection{Backup Overhead}

We evaluate the overhead of the backup operation in RevDedup.  The server
initially has no data.  We submit the 96 daily versions to the RevDedup
server in the order of their creation dates.  Recall from
\S\ref{subsec:implementation} that the server performs several steps: (i)
writing unique segments to disk and linking to existing segments, (ii)
building the index, (iii) searching for duplicates, and (iv) removing
duplicate blocks.  Steps~(ii)-(iv) correspond to the reverse deduplication
overhead, which we evaluate here.  In our measurements, we call {\tt sync()}
after writing unique segments to disk in Step~(i). 

\begin{figure}[t]
\centering
\begin{tabular}{cc}
\includegraphics[width=2.8in]{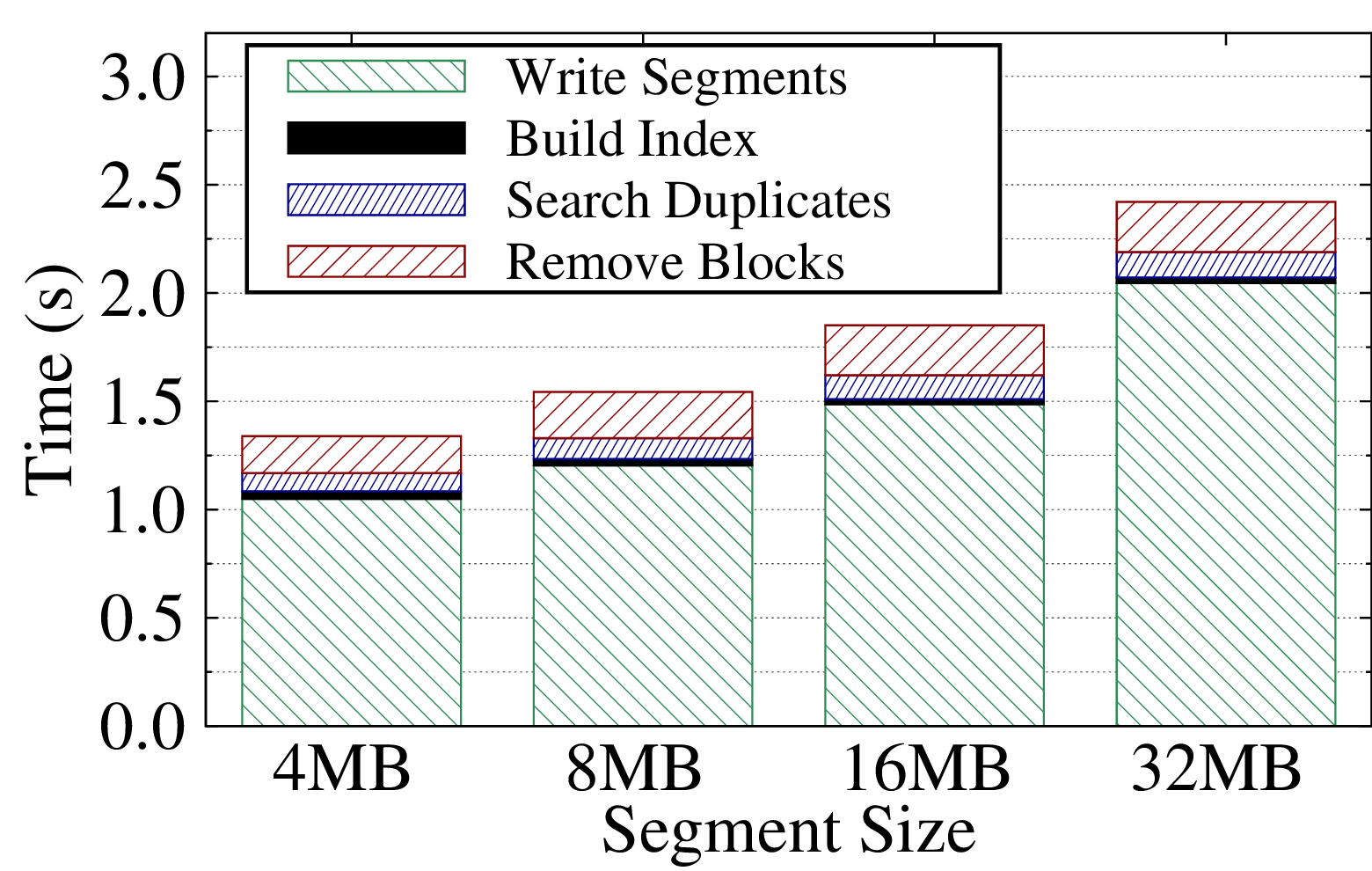} &
\includegraphics[width=2.8in]{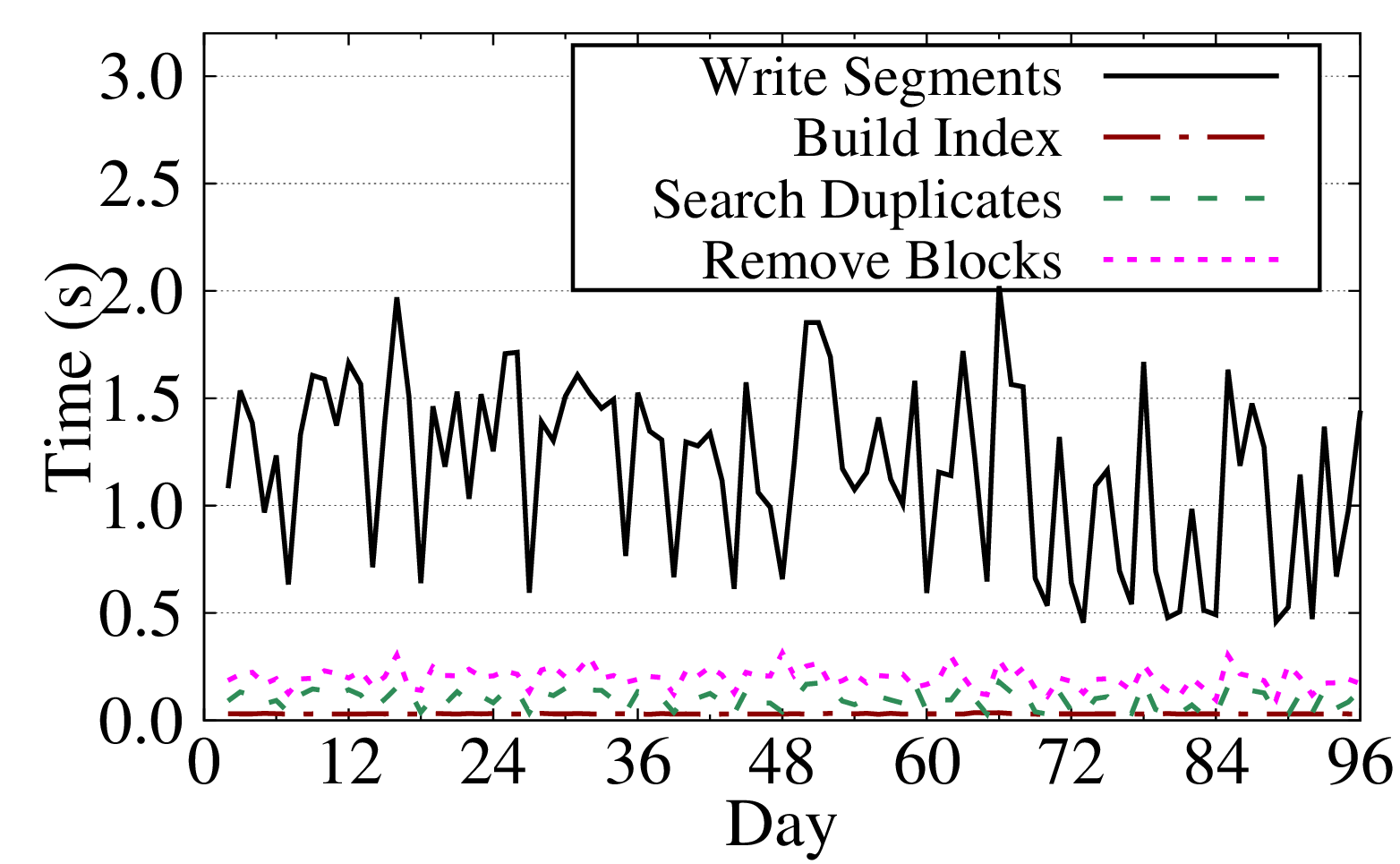} \\
\mbox{\small (a) Different segment sizes} &
\mbox{\small (b) Different days (segment size is 32MB)}
\end{tabular}
\caption{Backup time (per version) of RevDedup for a long-chained VM.}
\label{fig:synup}
\end{figure}

Figure~\ref{fig:synup}(a) shows the average time of writing a version, and
provides a breakdown for different steps.  Reverse deduplication 
(i.e., Steps~(ii)-(iv)) accounts for only 15-22\% of the total backup time.
Specifically, block removal contributes to most of the running time in reverse
deduplication, since it needs to copy data segments in segment compaction
(if the percentage of removed blocks is above 20\% as specified in our default
 setting).  Figure~\ref{fig:synup}(b) shows the instantaneous time of writing
each version, with the segment size fixed at 32MB.  The reverse deduplication
overhead remains small compared to the backup operation during the entire
period.  
%Also, the reverse deduplication process has fairly stable running time.  The
%reason is that our VM has a similar amount of daily changes, and hence the
%reverse deduplication process removes a similar amount of duplicates.  

\subsubsection{Block Removal Overhead}

\begin{figure}[t]
\centering
\begin{tabular}{cc}
\includegraphics[width=2.8in]{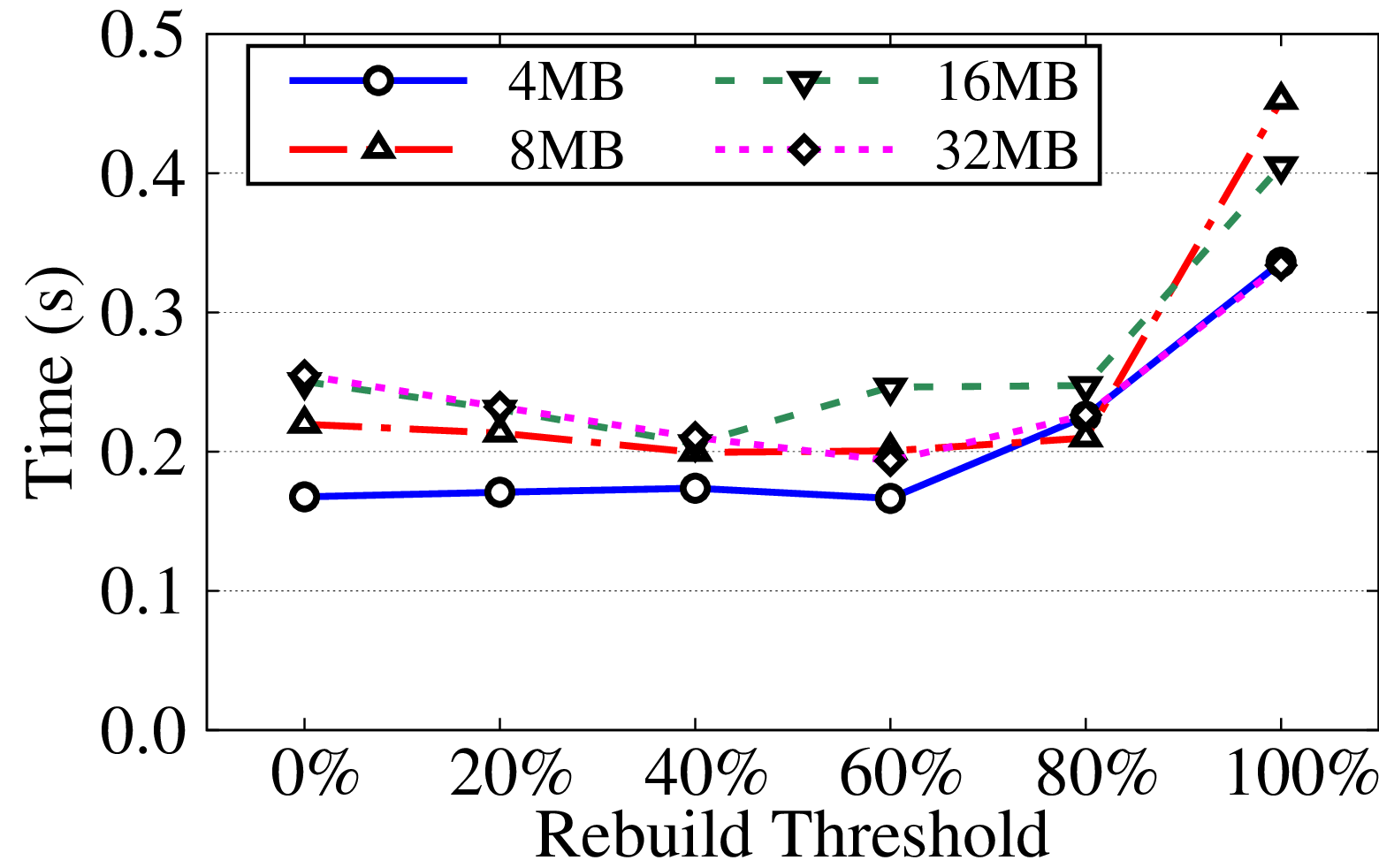} &
\includegraphics[width=2.8in]{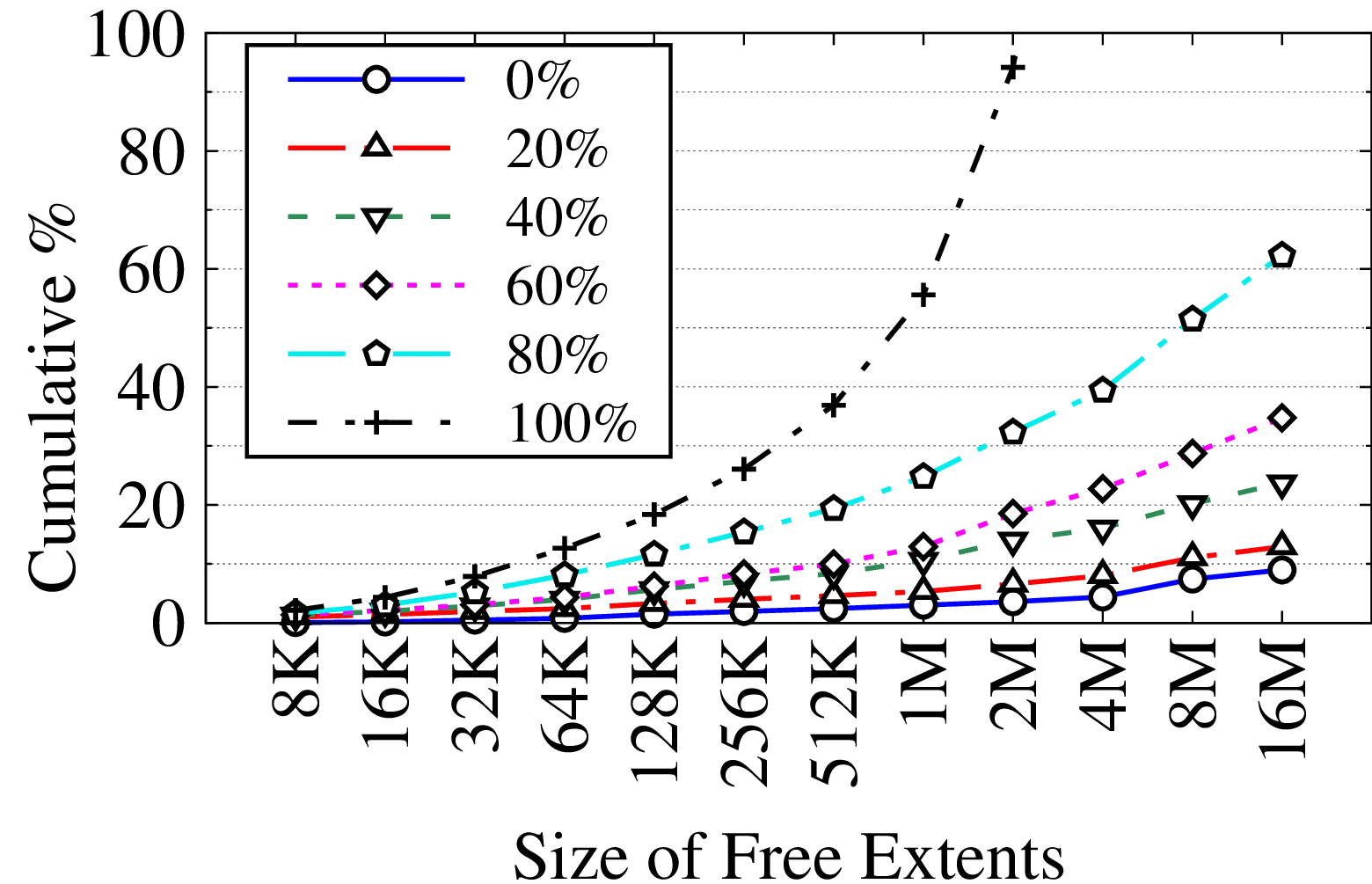} \\
\mbox{\small (a) Block removal time per version} &
\parbox[t]{3in}{\small (b) Disk fragmentation distribution with segment size
fixed at 32MB (we only plot data within 100\%)}
\end{tabular}
\caption{Effects of different rebuild thresholds in block removal.}
\label{fig:revdedup}
\end{figure}

We evaluate the impact of different rebuild thresholds on block removal
(see \S\ref{subsec:removal}).  We submit the 96 versions using different
thresholds.  Figure~\ref{fig:revdedup}(a) shows the
average time needed for RevDedup to perform block removal for each VM.
Initially, the block removal time decreases with the threshold, since RevDedup
uses block punching more frequently and incurs fewer I/Os of block content in
segment compaction.  However, the block removal time increases as the
threshold further increases, since segment compaction has less overhead with
fewer blocks being copied, while block punching removes more blocks and incurs
higher file system metadata overhead.  Overall, when the rebuild threshold is
less than 80\%, the block removal time is within 0.26s, which remains small
compared to the overall backup time (see Figure~\ref{fig:synup}). 

We also evaluate the impact on disk fragmentation for different rebuild
thresholds.  We use the Linux utility {\tt e2freefrag} to report the percentage
distribution of the sizes of free extents (i.e., the contiguous regions of
free blocks) in the whole file system partition after storing all 96 versions. 
We normalize the percentage for each size range of free extents by dividing
the total size of such free extents by the size of the actual data being
stored (after deduplication).  Here, we fix the segment size at 32MB.  
We say a free extent is small if its size is less than the 32MB segment size.
A high percentage of small free extents means that the disk is more
fragmented.  Figure~\ref{fig:revdedup}(b) shows the cumulative percentage
distribution.  A steeper curve
implies that the degree of disk fragmentation is higher. The figure shows
significant disk fragmentation when the threshold is at least 40\%.  When the
threshold reaches 100\% (i.e., block punching only), the cumulative percentage
goes beyond 100\%, meaning that the amount of small free extents is greater
than the amount of stored data.  
%Note that we still see a certain degree of disk fragmentation even when the
%threshold is 0\% (i.e., no block punching), since the underlying file system
%has a full control of the final placement of segments.

\subsubsection{Read Overhead}
\label{subsec:syntheticread}

\begin{figure}[t]
\centering
%\begin{tabular}{c@{\ }c}
%\hspace{-0.1in}
\includegraphics[width=2.8in]{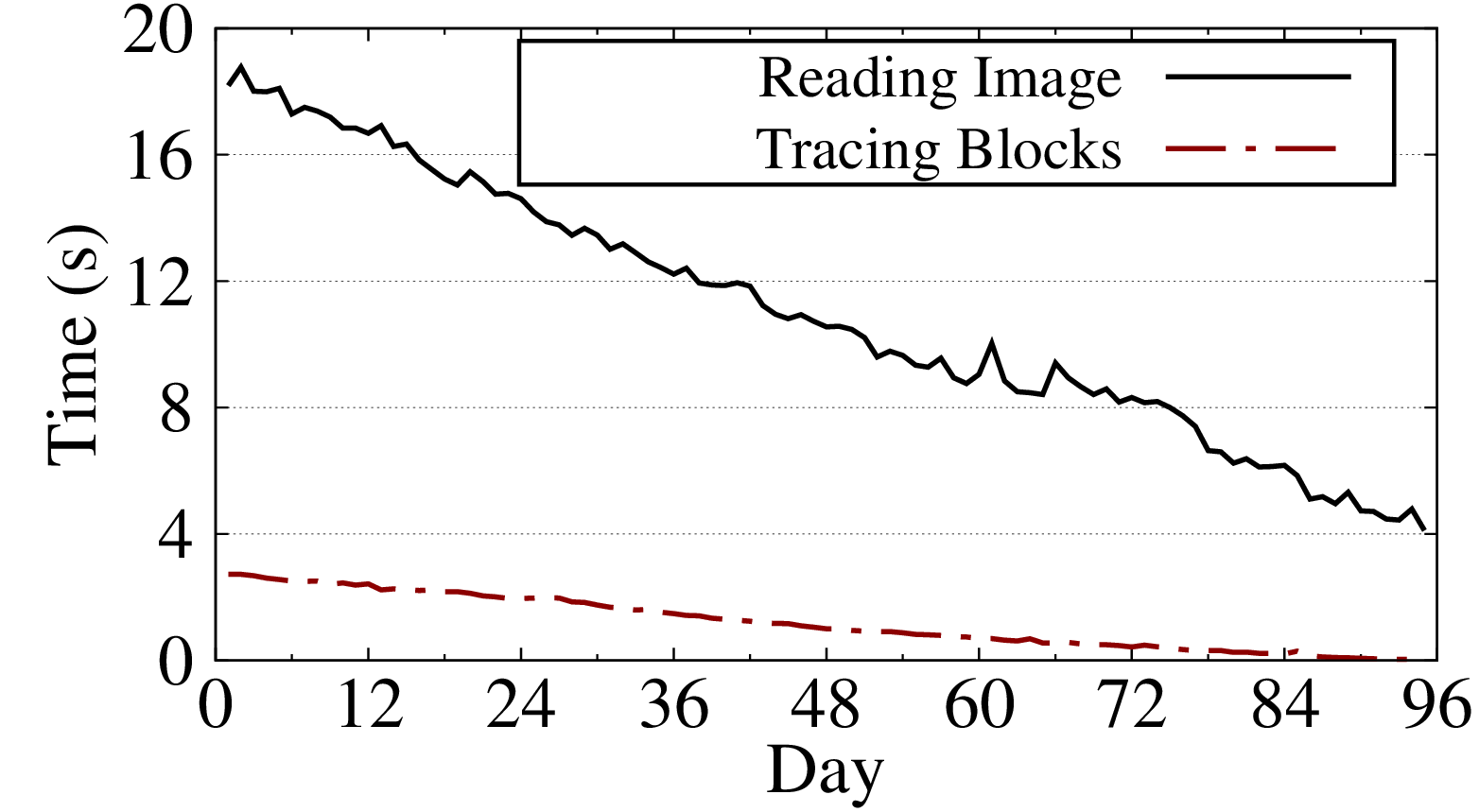}
%\includegraphics[width=1.5in]{charts/syn/downs} \\
%\mbox{\small (a) Tracing overhead} &
%\mbox{\small (b) Seek overhead}
%\end{tabular}
\vspace{-6pt}
\caption{Tracing overhead of RevDedup for a long-chained VM (segment size is
32MB).}
\label{fig:syndown}
\end{figure}

Recall that reading a deduplicated block of an old version is done by tracing
a chain of the indirect references. We now evaluate such tracing overhead.  
We submit all 96 daily versions in the order of their creation dates, and then
read each version after all submissions.  Here, we fix the segment size at
32MB.  Figure~\ref{fig:syndown} shows the overall time of reading each 5GB
version and the time spent in tracing the indirect references for each
version. We see that the tracing step only accounts for at most 15\% of the
overall reading time and has small overhead.

%We also measure the seek overhead of reading each version.  We say a seek
%occurs if the start address of the currently accessed block does not
%immediately follow the end address of the last access block.
%Figure~\ref{fig:syndown}(b) shows the number of seeks incurred in reading
%each version.  The number of seeks decreases for newer versions.  We see that
%reading the latest version (i.e., 96th) incurs 89\% fewer seeks than reading
%the first version.  This shows how RevDedup shifts the fragmentation overhead
%to earlier versions. 

\section{Conclusions}
\label{sec:conclusion}

We present RevDedup, a deduplication system designed for VM disk image backup
in virtualization environments. RevDedup has several design goals: high
storage efficiency, low memory usage, high backup performance, and high
restore performance for latest backups.  The core design component of RevDedup
is reverse deduplication, which removes duplicates of old backups and
mitigates fragmentation of latest backups.  We extensively evaluate our
RevDedup prototype using different workloads and validate our design goals. 

{\bf Availability.} The source code of our RevDedup prototype presented in
this paper is available for download at: {\bf
http://ansrlab.cse.cuhk.edu.hk/software/revdedup}.

\begin{footnotesize}
\bibliographystyle{plain}
\bibliography{paper}
\end{footnotesize}

\end{document}